%% file: ECDLP.tex
\documentclass{llncs}

\usepackage[l2tabu,orthodox]{nag}
\usepackage[utf8]{inputenc}
\usepackage{ifxetex}

\usepackage{microtype}
\usepackage{xcolor}
\definecolor{oxygenorange}{HTML}{FFDD00}
\usepackage[color=oxygenorange]{todonotes}

\def\isanonymous{0}

\usepackage{ifthen}
\newcommand{\anonymous}[2]{
\ifthenelse{\equal{\isanonymous}{1}}
{{#1}}
{{#2}}
}

\newif\iffullversion
\fullversiontrue
\usepackage{amsmath}
\usepackage{amssymb}
\usepackage{qcircuit}
\usepackage{algorithmic}
\usepackage{algorithm}
\usepackage{tikz}
\usepackage{subcaption}
\usepackage{xspace}
\usepackage{mathtools}
\usepackage{multirow}

\usepackage{booktabs}  %% tables
\usepackage{comment}
\usepackage{enumitem}
\iffullversion
\usepackage{a4wide}
\fi
\usepackage{url}

\ifxetex%
\else%
\usepackage{embedfile}
\fi%

\definecolor{linkcolor}{rgb}{0.65,0,0}
\definecolor{citecolor}{rgb}{0,0.65,0}
\definecolor{urlcolor}{rgb}{0,0,0.65}
\definecolor{darkergreen}{rgb}{0,0.75,0}
\definecolor{blueboxfg}{RGB}{32,48,192}
\definecolor{blueboxbg}{RGB}{240,248,255}
\usepackage[hidelinks, colorlinks = true,
linkcolor=linkcolor, urlcolor=urlcolor, citecolor=citecolor]{hyperref}

\input{macros}

\makeatletter
\renewcommand*{\@fnsymbol}[1]{\ensuremath{\ifcase#1\or *\or \dagger\or \ddagger\or
		\mathsection\or \mathparagraph\or \|\or **\or \dagger\dagger
		\or \ddagger\ddagger \else\@ctrerr\fi}}
\makeatother

\title{Improved quantum circuits for elliptic curve discrete logarithms}
\anonymous{
	\author{ }
	\institute{ }
	}{
	\author{ \
        Thomas H{\"a}ner\inst{1} \and
        Samuel Jaques \inst{2} \thanks{Partially supported by the University of Oxford Clarendon fund.}\thanks{Most of this work was done by Samuel Jaques, while he was an intern at Microsoft Research.} \and
		Michael Naehrig  \inst{3} \and
		Martin Roetteler \inst{1} \and
		Mathias Soeken \inst{1}
	}
	\institute{
		Microsoft Quantum, Redmond, WA, USA \\
		\iffullversion\email{\{thhaner,martinro,a-masoek\}@microsoft.com}\fi
		\and
		Department of Materials, University of Oxford, UK\\
		\iffullversion\email{samuel.jaques@materials.ox.ac.uk}\fi
		\and
		Microsoft Research, Redmond, WA, USA \\
		\iffullversion\email{mnaehrig@microsoft.com}\fi
	}
}
\begin{document}

\maketitle

\begin{abstract}
	We present improved quantum circuits for elliptic curve scalar multiplication, the most costly component in Shor's algorithm to compute discrete logarithms in elliptic curve groups. We optimize low-level components such as reversible integer and modular arithmetic through windowing techniques and more adaptive placement of uncomputing steps, and improve over previous quantum circuits for modular inversion by reformulating the binary Euclidean algorithm. Overall, we obtain an affine Weierstrass point addition circuit that has lower depth and uses fewer $T$ gates than previous circuits. While previous work mostly focuses on minimizing the total number of qubits, we present various trade-offs between different cost metrics including the number of qubits, circuit depth and $T$-gate count. Finally, we provide a full implementation of point addition in the \qsharp quantum programming language that allows unit tests and automatic quantum resource estimation for all components. 
	
\end{abstract}

\section{Introduction}
\input{introduction}

\section{Preliminaries}
\input{preliminaries}

\section{Components}\label{sec:components}
\input{components}

\section{Modular arithmetic}
\input{modular-arithmetic}

\section{Elliptic curves}\label{sec:elliptic_curves}
\input{elliptic-curves}

\section{Results}\label{sec:results}
\input{numerical-estimates}

\anonymous{}{
\iffullversion
\subsubsection{Acknowledgements.}
We thank Dan Bernstein, Martin Ekerå, Iggy van Hoof, and Tanja Lange for helpful suggestions about elliptic curve arithmetic. We thank Martin Albrecht for lending computing power to run resource estimates.
\fi
}

\iffullversion
\bibliographystyle{plain}
\else
\bibliographystyle{abbrv}
\fi
\bibliography{ECDLP}

%\newpage
\appendix
\input{appendix}

\end{document}

%% file: macros.tex
\newcommand\ket[1]{\left\vert #1\right\rangle}
\newcommand\floor[1]{\left\lfloor #1 \right\rfloor}
\newcommand\gtxt[1]{\texttt{#1}}
\newcommand\qOutput[3]{\draw[fill=#3] (#1,#2) +(0,-2) -- +(3,0) -- +(0,2) -- +(0,-2);}

\newcommand\F{\mathbb{F}}

\newcommand\style{\mathcal }

\newcommand\Z{\mathbb{Z}}

\newcommand\ceil[1]{\left\lceil #1 \right\rceil}

\newcommand\qsharp{Q\#\xspace}

\newcommand\ancilla[3]{
\draw (#1 + 3,#3) -- (#2-3,#3);
}

%% file: introduction.tex
% !TeX root = ./ECDLP.tex
Shor's algorithm \cite{shor,Sho97} solves the discrete logarithm problem for finite abelian groups with only polynomial cost. When run on a large-scale, fault-tolerant quantum computer, its variant for elliptic-curve groups could efficiently break elliptic curve cryptography with parameters that are widely used and far out of reach of current classical adversaries. 

Barring the efficient classical post-processing of the measured data, Shor's quantum algorithm consists of three steps: first, a superposition of exponents is created, then those exponents control the evaluation of a group exponentiation, and finally a quantum Fourier transform is applied to the exponent register, which is then measured. The group operations in the exponentiation must be computed in superposition and this is by far the most expensive step of the algorithm. Thus, the precise cost of Shor's algorithm depends on a detailed resource estimation for implementing the group operation on a quantum computer. For solving the elliptic curve discrete logarithm problem (ECDLP), the relevant operation is the repeated controlled addition of classical elliptic curve points to an accumulator point in a quantum register.

The first detailed discussion of the elliptic curve case was given by Proos and Zalka in \cite{pz2003}. Based on this work, Roetteler et al. \cite{rnsl2017} (hereinafter referred to as RNSL) presented explicit quantum circuits for point addition and all its components and automatically derive their resource estimates from a concrete implementation. Both papers focus on minimizing the number of qubits required to run the algorithm, since its polynomial runtime implies that it will run fast once an adversary has enough qubits to do so.
They count the required number of \emph{logical} qubits. For example, RNSL estimate that Shor's algorithm needs 2330 logical qubits to attack a 256-bit elliptic curve. Under plausible assumptions about physical error rates, this could translate into $6.77\cdot 10^{7}$ physical qubits \cite{gm2019}. But the number of logical qubits is not the only important cost metric, and one might prioritize others such as circuit depth, the total number of gates, or the total number of likely expensive gates such as the Toffoli gate or the $T$ gate. 

Our goal in this work is not only to improve the circuits proposed by RNSL~\cite{rnsl2017}, but also to explore different trade-offs favoring different cost metrics. To this end, we provide resource estimates for point addition circuits optimized for depth, $T$ gate count, and width, respectively.
We also report on initial experiments with automatic optimization for $T$-depth and $T$ gate count. By using the automatic compilation techniques presented in~\cite{msc+19}, we find low $T$-depth and low $T$-count circuits for a modular multiplication component and show significant improvements compared to their manually designed counterparts, however, at a very high cost to the number of qubits.

Beyond alternative choices for low-level arithmetic components, we also improve the higher-level structure of RNSL's circuit. While many components stay the same, the most dramatic improvements come from windowing techniques similar to those proposed by Gidney and Ekerå in \cite{ge2019} and a better memory management via \emph{pebbling}. For example, instead of copying out the result in an out-of-place circuit that uses Bennett's method for embedding an irreversible function in a reversible computation, the result can be used for the next operation before it is uncomputed. This technique does not treat modular operations merely as black boxes, but can adaptively reduce the cost of the higher-level circuit they are used in. Along with a reformulation of the binary extended Euclidean algorithm, it significantly reduces costs for the modular inversion circuit.

One of our main contributions is a modular, testable library\footnote{Our code will be released under an open source license.} of functions for elliptic curve arithmetic in the \qsharp programming language for quantum computing \cite{qsharp}. These incorporate different possible choices for  subroutines like addition and modular multiplication. Besides enabling unit testing for all components, the \qsharp development environment allows automated resource estimation. 

We strictly improve RNSL's estimates under all metrics. For example, for solving the ECDLP on a 256-bit elliptic curve, we reduce the number of qubits from $2338$ to $2124$, improve the $T$-count by a factor of 119 and the $T$-depth by a factor of 54. \iffullversion Asymptotically, we estimate that the number of $T$ gates in our circuit for Shor's algorithm scales as $436n^3 + o(n^3)$.

\fi
Under a different trade-off optimizing for depth, our circuit costs $2^{33}$ $T$ gates with $T$-depth of $2^{25}$ and $2871$ qubits. Compared to RNSL, this is a factor $6000$ reduction in $T$-depth with only a 22\% increase in width. \iffullversion Asymptotically, our circuits can achieve $1115n^3/\lg n + o(n^3/\lg n)$ $T$ gates, or $285n^2+o(n^2)$ $T$-depth.\fi

Extrapolating analogous values, breaking RSA-3072 would cost $2^{34}$ $T$ gates and $9287$ logical qubits \cite{ge2019}. This suggests that, at similar classical security levels, elliptic curve cryptography is less secure than RSA against a quantum attack.

%%% Local Variables:
%%% mode: latex
%%% TeX-master: "ECDLP"
%%% End:

%% file: preliminaries.tex
% !TeX root = ./ECDLP.tex
This section only gives a very brief discussion of the basic concepts used in this work. For a more detailed introduction to quantum computing, we refer to \cite{nc2000}, for Shor's algorithm see \cite{shor} and \cite{Sho97} and for its ECDLP variant \cite{pz2003} and \cite{rnsl2017}.

\subsubsection{Quantum computing.}
A quantum computer acts on \emph{quantum states} by applying quantum gates to its qubits. A quantum state is denoted by $\ket{x}$ for some label $x$. We work entirely with computational basis states, so $x$ is always a bit string. As the fundamental gate set we use the Clifford$+T$ gate set and assume that the $T$ gate is by far the most expensive, including measurements. This is a plausible assumption because, in a quantum computer using a surface code for error correction, $T$ gates consume special states which require many qubits and many surface code cycles to produce, and surface codes require frequent measurements for all gates \cite{fmmc2012}. A quantum algorithm is described by a sequence of gates in the form of a quantum circuit. We use standard quantum circuit diagrams\iffullversion, with black triangles representing output registers\fi. For circuit design and testing, we use circuits built from NOT, CNOT and Toffoli gates as those can be simulated efficiently at scale on classical inputs~\cite{rnsl2017}. However, for cost estimation, we use decompositions over the Clifford$+T$ gate set, such as the ones introduced in \cite{Amy13,gidney2018,selinger2013}. 

\subsubsection{Shor's algorithm for the ECDLP.}
Let an instance of the ECDLP be given by two $\F_p$-rational points $P,Q \in E(\F_p)$ on an elliptic curve $E$ over a finite field of large characteristic $p$ such that $\mathrm{ord}(P) = r$, $Q \in \langle P \rangle$. The problem is to find the unique integer $m\in\{1,\dots, r\}$ such that $Q=mP$. 
Shor's algorithm applies a Hadamard transform to two registers with $n+1$ qubits initialized to $\ket{0}$
to create the state $\frac{1}{2^{n+1}} \sum_{k,\ell=0}^{2^{n+1}-1} \ket{k}\ket{\ell}$.
Next, the state
$\frac{1}{2^{n+1}} \sum_{k,\ell=0}^{2^{n+1}-1} \ket{k}\ket{\ell}\ket{k P +\ell Q}$ is computed using the elliptic curve group law. 
A quantum Fourier transform $\mathrm{QFT}_{2^{n+1}}$ on $n+1$ qubits is applied to both $\ket{k}$ and $\ket{\ell}$ and the $2(n+1)$ qubits in $\ket{k}\ket{\ell}$ are measured. Classical post-processing then yields the discrete logarithm $m$. We assume that the algorithm is modified using the semiclassical Fourier transform method \cite{GN:96}, which means that a small number of qubits can be re-used to act as the $2n+2$ qubits for $\ket{k}\ket{\ell}$. RNSL use only one qubit for this \cite{rnsl2017}, but we use more than one to allow windowed arithmetic (see Section~\ref{sec:point_addition_windowing}). The most cost-intensive part is the double-scalar multiplication to compute $\ket{kP+\ell Q}$.

\subsubsection{Functions as quantum circuits.}
The elliptic curve group law is built from various classical functions that operate on bit strings of varying lengths $n$. For any function $f$, we use $U_f$ to denote a quantum circuit that computes $f$.

We often want $U_f$ to compute $f$ \emph{in-place}, meaning it has the action $U_f: \ket{x} \mapsto \ket{f(x)}$ on inputs $x\in \{0,1\}^n$. If $U_f$ is built out of Clifford$+T$ gates, each gate is easy to invert; thus, an in-place circuit $U_f$ automatically yields an in-place circuit $U_f^\dagger$ that computes $f^{-1}$. As quantum circuits need to be reversible, an in-place circuit is not always possible, e.g. if $f$ is not injective.

\iffullversion
As an example consider modular multiplication. Let $p$ be an integer modulus. The function $f: (x,y)\mapsto (x,xy\mod p)$ is injective for inputs $x,y \in \Z/p\Z$ that are co-prime to $p$. Its inverse is modular division. At present, the best circuits we know to compute modular division use some variant of the Euclidean algorithm. Thus, an in-place modular multiplication circuit would either be more expensive than the Euclidean algorithm or would provide a state-of-the-art circuit for modular division.
\fi

When an in-place circuit is not possible, $U_f$ needs to implement $f$ \emph{out-of-place} as $\ket{x}\ket{0}^n\mapsto\ket{x}\ket{f(x)}$. Some circuits might require a number $m$ of \emph{auxiliary qubits}, such that
\iffullversion
\begin{equation}
U_f:\ket{x}\ket{0}^n\ket{0}^m\xmapsto{U_f}\ket{x}\ket{f(x)}\ket{g(x)},
\end{equation}
\else
$U_f:\ket{x}\ket{0}^n\ket{0}^m\mapsto\ket{x}\ket{f(x)}\ket{g(x)}$,
\fi
where $g$ is some function of the input.
The auxiliary qubits are entangled with the other registers, and $g$ must be \emph{uncomputed} before the end of the computation to restore them to their original state. In our circuit diagrams, white triangles indicate such outputs.

If it is too costly to compute $g(x)$ from $x$ and $f(x)$, one can use a method due to Bennett~\cite{bennett1973} to clean any auxiliary qubits. By adding another $n$-qubit register, the output $f(x)$ is ``copied'' to that register using CNOT gates. Then, the inverse $U_f^\dagger$ is applied. This trick roughly doubles the cost to reversibly compute $f$.

In a sequence of out-of-place circuits, uncomputing early steps prevents us from uncomputing later steps. To make the full algorithm work, we either need to keep intermediate steps at the cost of an increasing number of qubits or recompute them repeatedly at the cost of additional gates. This is an instance of a general problem known as a \emph{pebbling game}. 

\subsubsection{Controlled circuits.}
As larger circuits are composed from smaller ones, the smaller circuit often needs to be controlled with a single qubit. We can ``promote'' each Clifford$+T$ gate in the smaller circuit into a controlled variant, which is expensive, e.g., for CNOT and Toffoli gates. Thus, we want to optimize which gates we control. For example, a circuit using Bennett's trick can be controlled by changing only the middle CNOT gates into Toffoli gates.
For other circuits, we design controlled versions as needed.

%%% Local Variables:
%%% mode: latex
%%% TeX-master: "ECDLP"
%%% End:

%% file: components.tex
% !TeX root = ./ECDLP.tex
\subsubsection{Design strategies.}
A full cost estimate of Shor's algorithm requires estimates at all layers of the architecture, including error correction, layout, and possibly architecture design. We focus only on the logical layer, and provide circuits that operate on abstracted logical qubits. From this level we cannot decide which design choices will be optimal. A shallower circuit with more logical qubits would have smaller error correction overhead and could have fewer physical qubits.

Thus, we provide different approaches and tradeoffs. We measure the $T$-depth, $T$-count, depth including all gates, and total number of qubits used (``width''). We focus on three strategies, favouring depth, $T$-count, or width. We could instead make different choices for each sub-circuit of Shor's algorithm, resulting in a large parameter space for potential optimization. Since we wrote \qsharp functions for all circuits, future work could combine different choices for each step.

\subsubsection{Toffoli and AND gates.}\label{seq:toffoli}
As explained by RNSL in the introduction of \cite{rnsl2017}, circuits that are expressed as Toffoli gate networks can be implemented exactly over the 
Clifford$+T$ gate set and can be classically simulated and tested. Therefore, our circuits are designed using the same approach.
The only source of $T$-gates in such a circuit is from Toffoli gates. In many instances, we know that the output qubit is in the $\ket{0}$ state and we can use a dedicated AND circuit with a lower $T$-count instead. We use an AND gate design which combines Jones' and Selinger's AND gates \cite{Jones13,selinger2013}.
It uses 4 $T$ gates, while a Toffoli gate uses 7, and the inverse AND$^\dagger$ uses no $T$ gates while Toffoli$^\dagger$ uses 7. AND gates use 1 auxiliary qubit for $T$-depth 1; Toffoli gates can use 4 auxiliary qubits for T-depth 1 or none for $T$-depth 3, see \cite{Amy13}.

\subsubsection{Integer addition.}
The adder of lowest known $T$-count is Gidney's modification of the Cuccaro et al. adder \cite{cdkm2004} (hereinafter called the CDKMG adder), which uses only $4n$ $T$ gates to add two $n$-bit numbers, but uses $n$ auxiliary qubits \cite{gidney2018}. The adder of lowest known $T$-depth is the carry lookahead adder of Draper et al. \cite{dkrs2004} (hereinafter the DKRS adder), with logarithmic depth but $2n-O(n)$ auxiliary qubits. The adder of lowest width is due to Takahashi et al. \cite{ttk2009} (hereinafter the TTK adder), which computes in-place using no auxiliary qubits. RNSL used the TTK adder. 
The DKRS adder uses $10n$ Toffoli gates, but of these, $4n$ can be replaced with an AND or AND$^\dagger$ gate.

\iffullversion
\subsubsection{Controlled addition.}
\fi
We provide new methods for controlling the addition circuits.
The DKRS adder uses a circuit to propagate carries, which is uncomputed. These gates do not need to be controlled. The remaining gates which must be promoted to controlled versions are all CNOT gates, leading to only a slight increase in the T-count.
To control the CDKMG adder, we promote two CNOT gates per bit to Toffoli gates, as Figure \ref{fig:controlled-gidney-add} shows, and we change the final CNOT for the carry qubit to a Toffoli gate. 
Unfortunately, for both addition circuits, the new Toffoli gates cannot be replaced with AND gates.

\iffullversion
\begin{figure}
\centering
\input{quantum-circuits/Qpic/gidney-control}
\caption{Controlled addition block, adapted from \cite{gidney2018}.}\label{fig:controlled-gidney-add}
\end{figure}
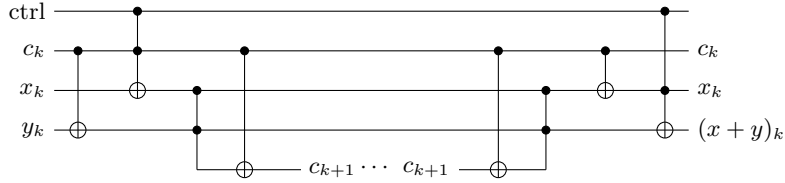
\fi

\subsubsection{Constant addition.}
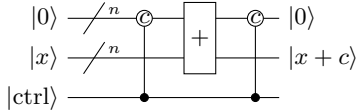
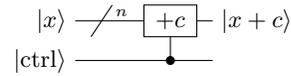
\begin{figure}
\centering
\begin{subfigure}[b]{0.4\textwidth}
\centering
\input{quantum-circuits/Qpic/control-copy-add}
\caption{Copying the classical constant $c$ is controlled, followed by uncontrolled addition.}\label{fig:ctrladdcnst}
\end{subfigure}
\hfill
\begin{subfigure}[b]{0.4\textwidth}
\centering
\input{quantum-circuits/Qpic/control-curry-add}
\caption{The addition circuit is curried for the constant $c$, then the full circuit is controlled.}\label{fig:ctrladdcurry}
\end{subfigure}
\caption{Two methods for controlling addition by a constant $c$.}\label{fig:constant-controlled-add}
\end{figure}

When tackling an ECDLP instance, the prime modulus is a classically known integer, and we need a circuit to add it to integers encoded in quantum registers. The simplest method allocates a new quantum register, inputs the integer into the quantum register, runs any quantum addition circuit, then uncomputes the integer from the quantum register to free the qubits. 
This simple method is easy to control. Only copying the integer into the quantum register is controlled, and then an uncontrolled addition is used, as shown in Figure~\ref{fig:ctrladdcnst}. Copying an integer uses only $X$ gates, so a controlled copy operation only uses CNOT gates, giving the same $T$-cost as uncontrolled quantum addition. We use this strategy with the CDKMG and TTK adders.

An alternative is to ``curry'' the quantum addition circuit. In the DKRS adder, the qubits of one of the two inputs are only used as controls for CNOT and Toffoli gates. We can replace these with $X$ and CNOT gates which are conditionally applied according to the bits of the classical integer. A controlled classical addition with this method needs to control the entire curried circuit (see Figure~\ref{fig:ctrladdcurry}), but we found that this is the most efficient approach with the DKRS adder.

\subsubsection{Comparing integers.}
An addition circuit immediately gives a comparator by the one's complement trick. For integers $x$ and $y$ represented in binary, if we let $x'$ denote the one's complement (all bits flipped), then $(x'+y)' = x - y$. If $y > x$, then $x - y < 0$ and the leading bit of $(x'+y)'$ will be 1. Thus, to construct a comparator, we simply flip all bits in $x$, compute the first half of an addition circuit, copy out the carry, then uncompute the addition circuit. With a control, we only need to control the final copy of the carry bit. 
We use this technique for comparators based on the previous adders. DKRS provide a comparator which we did not use for ease of implementation. 

\subsubsection{Fan-out and fan-in of control qubits.}
When a single qubit controls multiple parallel gates, they must be performed sequentially. To avoid this increase in depth, we opt to allocate extra qubits and fan-out the control \cite{moore1999} to these qubits using CNOTs. Then these qubits can control all the gates in parallel before we clear them again. For $n$ simultaneous controlled gates, this requires $n$ auxiliary qubits and at most $4n$ CNOT gates in depth $\ceil{\lg n}$, but no T gates. Our low-width optimization does not do this.

The fanned-out auxiliary control qubits could be retained to control many gates, depending on the application. Because the \qsharp language allocates qubits in a stack, this is often difficult. A function that allocates clean auxiliary qubits must restore them to the $\ket{0}$ state before returning. Because of this difficulty and the low gate cost of fan-out, we do not make such optimizations.

To control a single gate with the logical AND of $n$ qubits requires a fan-in of control qubits. For low width, we use Barrenco et al.'s method \cite{barencoetal1995}, as RNSL did. This performs a multi-AND in-place but with $8n$ Toffoli gates and linear depth. If we instead allocate $n-1$ auxiliary qubits and ``compress'' the controls with a tree of AND gates, it requires only $n-1$ AND gates and AND-depth $\ceil{\lg n}$.

%%% Local Variables:
%%% mode: latex
%%% TeX-master: "ECDLP"
%%% End:

%% file: quantum-circuits/Qpic/gidney-control
PREAMBLE \providecommand{\ket}[1]{\left\vert #1\right\rangle}

ctrl W \textrm{ctrl}
ck W c_k c_k
ik W x_k x_k
tk W y_k (x+y)_k
ck1 W {} {c_{k+1} \cdots} c_{k+1}

ck +tk
ctrl ck +ik

ck1 START

ik tk +ck1 shape=0

ck +ck1

ck1 END

+ik shape=0 
TOUCH
+ik shape=0
+ik shape=0

ck1 START

ck +ck1

ik tk +ck1 shape=0
ck1 END
ck +ik
ctrl ik +tk

%% file: quantum-circuits/Qpic/control-copy-add
PREAMBLE \providecommand{\ket}[1]{\left\vert #1\right\rangle}

y W \ket{0} \ket{0}
x W \ket{x} \ket{x + c}
ctrl W \ket{\textrm{ctrl}}

y / n
x / n

ctrl +y op=$c$

y x G $+$ 

ctrl +y op=$c$

%% file: quantum-circuits/Qpic/control-curry-add
PREAMBLE \providecommand{\ket}[1]{\left\vert #1\right\rangle}

x W \ket{x} \ket{x + c}
ctrl W \ket{\textrm{ctrl}}

x / n

x G width=20 $+c$ ctrl

%% file: modular-arithmetic.tex
% !TeX root = ./ECDLP.tex
Because modular reduction is irreversible, we cannot design an in-place circuit which maps $x$ to $x\bmod N$ without creating some auxiliary qubits that represent the quotient $\lfloor x/N\rfloor$. Any algorithm for modular arithmetic that uses many modular reductions thus creates significant qubit overhead. Instead, we design bespoke circuits for each operation. Primarily, we follow RNSL \cite{rnsl2017}.
We find that working in Montgomery representation \cite{mont85} has the lowest costs. For an odd modulus $p$ with $n=\ceil{\lg p}$, the Montgomery representation of an integer $x$ is $x 2^n \bmod p$. 

We use the modular addition circuit from RNSL. Note that addition in Montgomery representation is the same as in standard representation. For controlled modular addition, we only need to control two operations, as Figure \ref{fig:controlled-modular-addition} shows. This automatically gives us modular addition by a constant, by replacing the quantum-quantum addition and comparison circuits by their quantum-classical counterparts. 

\iffullversion
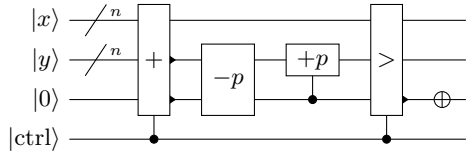
\begin{figure}
\centering
\input{quantum-circuits/Qpic/control-modular-add}
\caption{Controlled modular addition.}\label{fig:controlled-modular-addition}
\end{figure}
\else
\begin{figure}[h]
\centering
\begin{subfigure}{0.5\textwidth}
\resizebox{1.1\textwidth}{!}{
\input{quantum-circuits/Qpic/gidney-control}
}
\caption{Addition block from \cite{gidney2018}, with controls.}\label{fig:controlled-gidney-add}
\end{subfigure}
\hfill
\begin{subfigure}{0.4\textwidth}
\resizebox{\textwidth}{!}{
\input{quantum-circuits/Qpic/control-modular-add}
}
\caption{Controlled modular addition.}\label{fig:controlled-modular-addition}
\end{subfigure}
\caption{Efficient controlled addition.}\label{fig:controlled-addition}
\end{figure}
\fi

\subsection{Multiplication and squaring}
RNSL provide two circuits each for multiplication and squaring. We use the one that operates on $n$-bit integers in Montgomery form with $2n$ additions and $n$ halvings, and justify this choice in Appendix~\ref{app:multiplication_variants}. This circuit is a direct translation of classical Montgomery multiplication.

\subsubsection{Windowed arithmetic.}\label{sec:multiplication_windowing}
Windowed multiplication, such as \cite{gidney2019}, is not directly possible in our setting because we multiply two quantum integers, but we can adapt some of these techniques to our setting. When computing $x\cdot y$, the Montgomery multiplier uses the $i$th bit of $\ket{x}$ to control an addition of $\ket{y}$ to the output register. It then copies the lowest bit of the output to an auxiliary register $\ket{m}$ and uses this bit to control addition of the modulus $p$. This ensures that the sum is even, and so we rotate the register to divide by 2. 

When using a window of size $k$, the integer $x$ is split into $k$-bit words $x_{(1)},\dots, x_{(n/k)}$, analogous to classical interleaved radix-$2^k$ Montgomery multiplication. The $k$-bit value $x_{(i)}$ is multiplied with $y$, adding the $(n+k)$-bit result to the output register. To add a suitable multiple of $p$ to the output to set the $k$ least significant bits to 0, these $k$ bits are copied to an auxiliary register $m_i$. The multiple $t_{m_i}p$ such that $t_{m_i}p + m_i \equiv 0\mod 2^k$ can be looked up from a classically pre-computed table $T$, where $T[m_i] = t_{m_i}p$ for $t_{m_i}\equiv p^{-1}m_i\mod 2^k$.

We use the bits in $m_i$ as an address for a sequential quantum look-up \cite{babbushetal2018}\iffullversion\footnote{This technique is referred to as \emph{QROM} in \cite{babbushetal2018}; some papers also call it \emph{QRAM}; we use \emph{quantum look-up} to distinguish it from quantum random oracles and to emphasize that it is constructed out of Clifford+$T$ gates and does not require special QRAM technology.}\fi, writing the resulting $(n+k)$-bit integer $t_{m_i}p$ into an auxiliary register. Then an uncontrolled addition of $t_{m_i}p$ into the output register clears the bottom $k$ bits, so we can cyclically rotate by $k$ bits. 
Figure~\ref{fig:windowed-multiplication} illustrates this process. 

\iffullversion
\begin{figure}[h]
\centering
\input{quantum-circuits/Qpic/mul-windowed}
\caption{Circuit for a single window of windowed add-and-halve multiplication. The $\texttt{cpy}$ gate copies $k$ bits with $k$ CNOT gates. The gate $\texttt{In}_{m_i}$ performs a quantum table look-up of $T[m_i]$, where $T$ is the table described in the text. The circuit in this figure is repeated $\ceil{n/k}$ times, with a final modular correction step, to perform a single modular multiplication.}
\label{fig:windowed-multiplication}
\end{figure}
\else
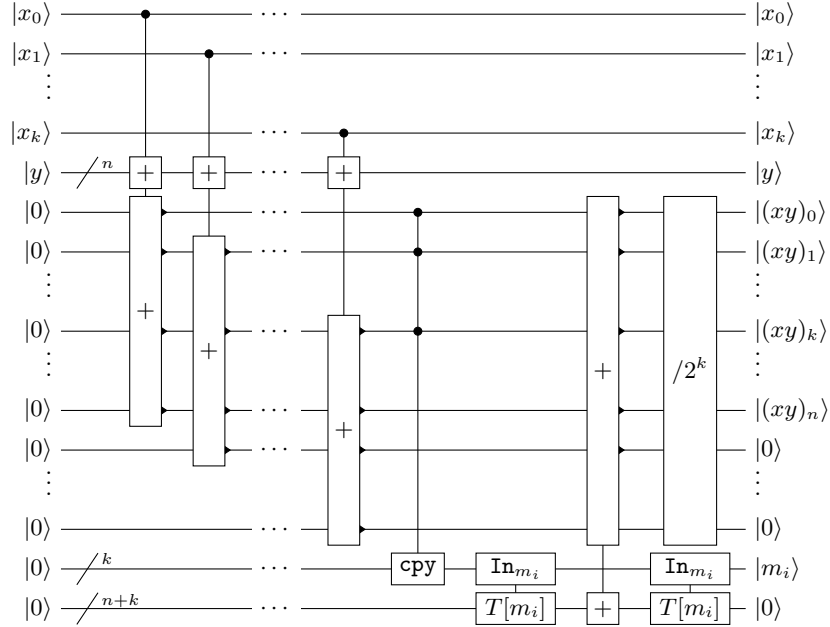
\begin{figure}[h]
\centering
\resizebox{!}{0.5\textwidth}{
\input{quantum-circuits/Qpic/mul-windowed}
}
\caption{Circuit for a single window of windowed add-and-halve multiplication. The $\texttt{cpy}$ gate copies $k$ bits with $k$ CNOT gates. The gate $\texttt{In}_{m_i}$ performs a quantum table look-up of $T[m_i]$, where $T$ is the table described in the text. The circuit in this figure is repeated $\ceil{n/k}$ times, with a final modular correction step, to perform a single modular multiplication.}
\label{fig:windowed-multiplication}
\end{figure}
\fi

\subsubsection{Pebbling.}\label{sec:multiplication_pebbling}
Given integers $x$, $y$, and $z$, one method to compute $xy+z$ in place is to first compute $xy$ in an auxiliary register, then to add it to the register containing $z$, and finally to uncompute $xy$. This works for any generic multiplication circuit, but at the cost of two multiplications. 

The circuit we use requires the Bennett method, and we replace its CNOT step with an addition. This is just a pebbling technique \cite{bennett1989}; we are keeping the auxiliary qubits until we have finished the next computation (an addition) before uncomputing them. The cost of this multiply-then-add is just the sum of the costs of a multiplication and an addition, rather than twice the cost of a multiplication and an addition.

\begin{figure}
\centering
\begin{subfigure}{0.45\textwidth}
\centering
\resizebox{1.1\textwidth}{!}{
\input{quantum-circuits/Qpic/QDLP-square-expanded}
}
\caption{Original circuit \cite{rnsl2017} with \gtxt{squ} circuit expanded, \gtxt{squ}$_A$ produces dirty auxiliary qubits in $\ket{t_1}$.}\label{fig:RNSL_square_expanded}
\end{subfigure}
\hfill
\begin{subfigure}{0.45\textwidth}
\centering
\resizebox{\textwidth}{!}{
\input{quantum-circuits/Qpic/QDLP-square-efficient}
}
\caption{More efficient version.}\label{fig:square_efficient}
\end{subfigure}
\caption{Improvement to the squaring circuit from \cite{rnsl2017} in the context of elliptic curve point addition.}\label{fig:squaring}
\end{figure}
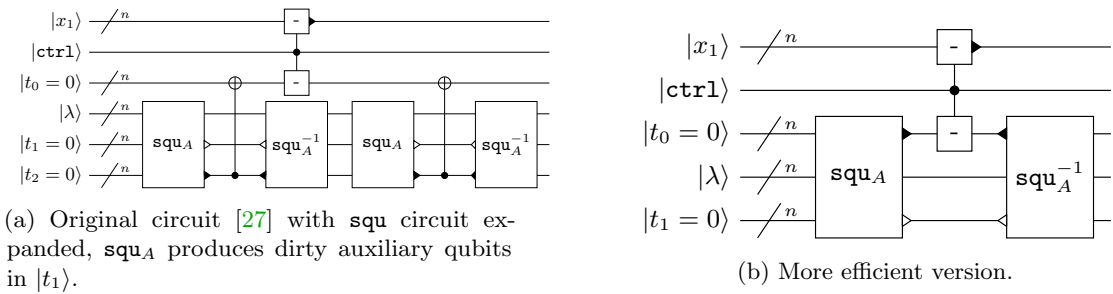

In particular, RNSL treat their squaring circuit as a black box in the full elliptic curve addition circuit. The computed square is only needed for one subtraction before it must be uncomputed; therefore we can use this multiply-then-add trick. Figure \ref{fig:squaring} shows how this saves almost half the gate cost and depth of the squaring, as well as saving an auxiliary register. 

\subsubsection{Automatic optimization.}
The multiply-then-add operation is a very costly component in the point addition circuit. We explore what is possible when reducing its $T$-depth and $T$-gate count. Appendix~\ref{app:autocompile} shows the results for a compilation using the method from~\cite{msc+19} after automatic optimization~\cite{tsam19} of a logic network generated from the operation. We achieve extremely low $T$-depth and $T$-count, but the number of logical qubits increases significantly. We leave further exploration of such techniques for future work.

\subsection{Modular inversion}
Modular inversion uses variants of the extended Euclidean algorithm (EEA). The EEA repeatedly divides integers. For two $n$-bit integers, we might expect each division requires $O(n^2)$ operations and, in the worst case, we need $O(n)$ divisions. In practice the complexity is smaller, because the complexity of the divisions becomes smaller in the course of the algorithm. Unfortunately, to exploit this fact, the circuit must be \emph{dynamic}: The number of divisions and the circuit for each division depend on the inputs. We cannot build such a quantum circuit, since we cannot observe the input (it may be in superposition).

We considered several approaches, but found that RNSL's circuit is the best. Appendix~\ref{app:inversion_variants} details our reasoning. However, we found several improvements to it.
Their circuit models an algorithm of Kaliski \cite{kaliski1995}, which applies a round operation (Algorithm \ref{alg:kaliski-round}) for up to $2n$ iterations. Conditional logic selects one of four different cases that can occur in each round. As a quantum circuit, this requires applying all four possible rounds, each with a different control. Algorithm \ref{alg:improved-kaliski-round} shows our alternative formulation based on swaps.

Figure \ref{fig:inversion_round} shows the round operation of RNSL's quantum circuit implementation of Kaliski's algorithm. This circuit repeats the controlled additions, doublings, and halvings, with different registers playing the role of input or output. Our method performs each of these operations once, using controlled swaps to arrange inputs and outputs for the respective case. The lower auxiliary bit and $\ket{m_i}$ uniquely specify the round and we use these bits to control a swap. This leads to Figure \ref{fig:improved_inversion_round}. A controlled $n$-bit swap is approximately the same cost as a controlled $n$-bit cyclic shift. 

\iffullversion
\begin{figure}
\centering
\begin{subfigure}{\textwidth}
\centering
\resizebox{\textwidth}{!}{
\input{quantum-circuits/Qpic/kaliski}}
\caption{Round operation from \cite{rnsl2017}.}\label{fig:inversion_round}
\end{subfigure}

\begin{subfigure}{\textwidth}
\centering
\resizebox{\textwidth}{!}{
\input{quantum-circuits/Qpic/kaliski-efficient}
}
\caption{Improved round operation.}\label{fig:improved_inversion_round}
\end{subfigure}
\caption{Improvement to the round operation in the binary extended Euclidean algorithm from \cite{rnsl2017} addressing the different cases by controlled swaps.}\label{fig:inversion_rounds}
\end{figure}
\else
\begin{figure}
    \centering
    \begin{subfigure}{0.49\textwidth}
    \centering
    \resizebox{\textwidth}{!}{
    \input{quantum-circuits/Qpic/kaliski}}
    \caption{Round operation from \cite{rnsl2017}.}\label{fig:inversion_round}
    \end{subfigure}
    \hfill
    \begin{subfigure}{0.49\textwidth}
    \centering
    \resizebox{\textwidth}{!}{
    \input{quantum-circuits/Qpic/kaliski-efficient}
    }
    \caption{Improved round operation.}\label{fig:improved_inversion_round}
    \end{subfigure}
    \caption{Improvement to the round operation in the binary extended Euclidean algorithm from \cite{rnsl2017} addressing the different cases by controlled swaps.}\label{fig:inversion_rounds}
\end{figure}
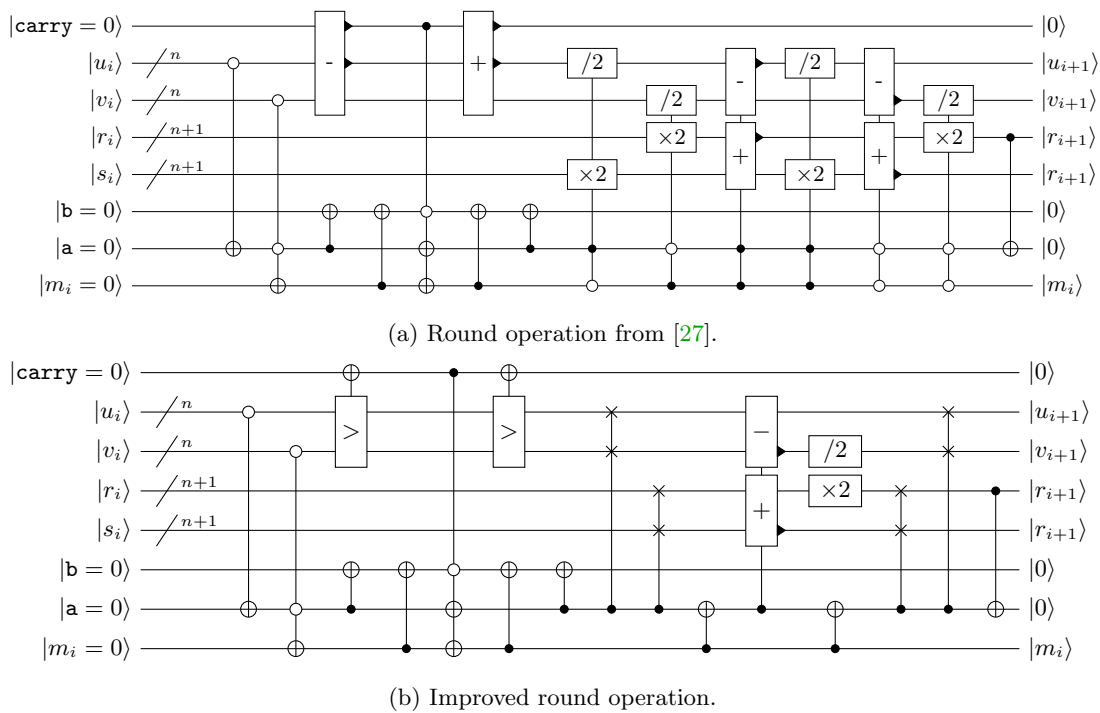
\fi

\subsubsection{Correcting pseudo-inverses in parallel.}
Classically, the Kaliski algorithm uses only $k$ rounds, with $n\leq k\leq 2n$ for $n$-bit integers. RNSL's quantum circuit executes $2n$ rounds, controlled in such a way that only $k$ actually modify the input. This produces an auxiliary qubit counter with a value of $2n-k$ and a pseudo-inverse output of $x^{-1}2^{-n+k}\bmod p$ for an input $x$. We want to output the Montgomery inverse $x^{-1}2^n\bmod p$, which requires correcting the pseudo-inverse. 

Instead of a separate doubling circuit like the one RNSL use, we add a doubling operation to each division round. After copying out the pseudo-inverse, during the subsequent uncomputation we use the same control that the round operation uses: in each of the $2n$ rounds, either we double the output register or perform a division round. These can be done in parallel, which improves the depth without increasing the total gate count. The result is that we compute $2n-k$ doublings, exactly what is needed to correct the pseudo-inverse.

\begin{figure}
\centering
\begin{subfigure}{0.4\textwidth}
\centering
\noindent\rule{\textwidth}{1pt}
\caption{Kaliski's algorithm}\label{alg:kaliski-round}
\noindent\rule{\textwidth}{1pt}
\begin{algorithmic}[1]
\IF{$u$ odd and $v$ even}
\STATE $v\leftarrow v/2$
\STATE$r\leftarrow 2r$
\ELSIF{$u$ even and $v$ odd}
\STATE $u\leftarrow u/2$
\STATE $s\leftarrow 2s$
\ELSIF{$u$ odd and $v$ odd and $u>v$}
\STATE $u\leftarrow (u-v)/2$
\STATE $r\leftarrow r+s$
\STATE  $s\leftarrow 2s$
\ELSIF{$u$ odd and $v$ odd and $v\geq u$}
\STATE $v\leftarrow (v-u)/2$
\STATE $s\leftarrow r+s$
\STATE $r\leftarrow 2r$
\ENDIF
\end{algorithmic}
\noindent\rule{\textwidth}{1pt}
\end{subfigure}
\hfill
\begin{subfigure}{0.4\textwidth}
\centering
\noindent\rule{\textwidth}{1pt}
\caption{Equivalent formulation}\label{alg:improved-kaliski-round}
\noindent\rule{\textwidth}{1pt}
\begin{algorithmic}[1]
\STATE $b_{swap}\leftarrow\gtxt{false}$
\IF{$u$ even and $v$ odd, or $u$ and $v$ both odd and $u>v$}
\STATE \textbf{swap} $u$ and $v$
\STATE \textbf{swap} $r$ and $s$
\STATE $b_{swap}\leftarrow\gtxt{true}$
\ENDIF
\IF{$u$ odd and $v$ odd}
\STATE $v\leftarrow v-u$
\STATE $s\leftarrow r+s$
\ENDIF
\STATE $v\leftarrow v/2$
\STATE $r\leftarrow 2r$
\IF{$b_{swap}$}
\STATE \textbf{swap} $u$ and $v$
\STATE \textbf{swap} $r$ and $s$
\ENDIF
\end{algorithmic}
\noindent\rule{\textwidth}{1pt}
\end{subfigure}
\caption{Kaliski's algorithm, and an equivalent formulation based on swaps.}\label{fig:kaliski-algorithm}
\end{figure}

\subsubsection{Modular division.}
\begin{figure}
\centering
\begin{subfigure}{0.33\textwidth}
\resizebox{\textwidth}{!}{
\input{quantum-circuits/Qpic/ECC-div}
}
\caption{Inversion steps in elliptic curve point addition from \cite{rnsl2017}. Auxiliary qubits are not shown. Each \gtxt{inv} block represents \gtxt{inv}$_A$, copying out, then \gtxt{inv}$_A^{-1}$ (with the same pattern for \gtxt{mul}). Thus, this circuit contains 4 \gtxt{inv}$_A$ steps and 2 $\gtxt{mul}_A$ steps. }\label{fig:RNSL_inversion}
\end{subfigure}
\hfill
\begin{subfigure}{0.62\textwidth}
\resizebox{\textwidth}{!}{
\input{quantum-circuits/Qpic/ECC-div-expanded}
}
\caption{More efficient modular division. $\scriptstyle{0}\vdash$ depicts initializing an auxiliary qubit to zero, and $\scriptstyle{\dashv 0}$ depicts removing an auxiliary qubit known to be zero. The doubling of the output $\lambda$ in the last step corrects the pseudo-inverse output.}\label{fig:div-efficient}
\end{subfigure}
\caption{Improvement to the modular division circuit from \cite{rnsl2017} by clearing and re-using auxiliary qubits before the full uncomputation.}
\end{figure}
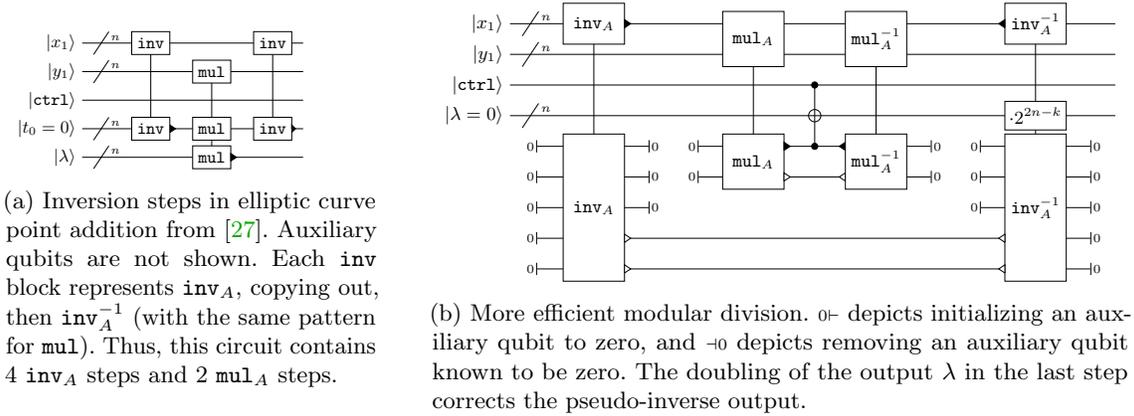

In elliptic curve point addition, the inversion is only necessary to compute a division. To divide an integer $x$ by $y$ modulo a prime $p$, RNSL first invert $y$ in an auxiliary register, perform the multiplication by the result, then invert $y$ again to uncompute the auxiliary qubits as shown in Figure~\ref{fig:RNSL_inversion}. 

Because the inversion uses Bennett's method, we pebble in the same way as the multiply-then-add circuit. To save qubits, we notice that after the inversion, three registers contain known values of $0$, $1$, and the modulus $p$. We can clear these auxiliary qubits with at most $3n$ parallel $X$ gates, then re-use them for modular multiplication. Denoting the inverse and multiplication operations with auxiliary qubits by $\gtxt{inv}_A$ and $\gtxt{mul}_A$, respectively, Figure~\ref{fig:div-efficient} depicts a full modular division. The pseudo-inverse correction is compatible with modular division.

%%% Local Variables:
%%% mode: latex
%%% TeX-master: "ECDLP"
%%% End:

%% file: quantum-circuits/Qpic/control-modular-add
PREAMBLE \providecommand{\ket}[1]{\left\vert #1\right\rangle}

x W \ket{x}
y W \ket{y}
z W \ket{0}
ctrl W \ket{ctrl}

x / n
y / n

x y z G $+$ ctrl
y z G $-p$ width=20
y G $+p$ width=20 z
x y z G $>$ \textrm{ctrl}
+z

%% file: quantum-circuits/Qpic/mul-windowed
PREAMBLE \providecommand{\ket}[1]{\left\vert #1\right\rangle}
PREAMBLE \providecommand{\gtxt}[1]{\texttt{#1}}

x0 W \ket{x_0} \ket{x_0}
x1 W \ket{x_1} \ket{x_1}
...midx1 W
xk W \ket{x_k} \ket{x_k}
y W \ket{y} \ket{y}

o0 W \ket{0} \ket{(xy)_0}
o1 W \ket{0} \ket{(xy)_1}
...omid1 W
ok W \ket{0} \ket{(xy)_k}
...omid2 W
on W \ket{0} \ket{(xy)_n}
on1 W \ket{0} \ket{0}
...omid3 W
onk W \ket{0} \ket{0}
m W \ket{0} \ket{m_i}
pt W \ket{0} \ket{0}

y / n
pt / n+k

y G $+$ o0 o1 ok on G $+$ x0 
y G $+$ o1 ok on on1 G $+$ x1

LABEL ...

y G $+$ ok on on1 onk G $+$ xk 

m G $\gtxt{cpy}$ o0 o1 ok width=20

m G $\gtxt{In}_{m_i}$ pt G $T[m_i]$ width=30

pt G $+$ o0 o1 ok on on1 onk G $+$ 

m G $\gtxt{In}_{m_i}$ pt G $T[m_i]$ width=30

o0 o1 ok on on1 onk G $/2^k$ width=20

%% file: quantum-circuits/Qpic/QDLP-square-expanded
PREAMBLE \providecommand{\ket}[1]{\left\vert #1\right\rangle}
PREAMBLE \providecommand{\gtxt}[1]{\texttt{#1}}

x1 W \ket{x_1}
y1 W \ket{y_1}
ctrl W \ket{\gtxt{ctrl}}
t0 W \ket{t_0=0}
m W \ket{\lambda}
t1 W \ket{t_1=0}
t2 W \ket{t_2=0}

x1 / n
y1 / n
t0 / n
m / n
t1 / n
t2 / n

m t1 t2 G:width=30 \gtxt{squ}$_A$
t2 +t0
m t1 t2 G:width=30 \gtxt{squ}$_A^{-1}$
t0 G {-} x1 G {-} ctrl
x1 G:width=30 $+3x^2$ ctrl
m t1 t2 G:width=30 \gtxt{squ}$_A$
t2 +t0
m t1 t2 G:width=30 \gtxt{squ}$_A^{-1}$

%% file: quantum-circuits/Qpic/QDLP-square-efficient
PREAMBLE \providecommand{\ket}[1]{\left\vert #1\right\rangle}
PREAMBLE \providecommand{\gtxt}[1]{\texttt{#1}}

x1 W \ket{x_1}
y1 W \ket{y_1}
ctrl W \ket{\gtxt{ctrl}}
t0 W \ket{t_0=0}
m W \ket{\lambda}
t1 W \ket{t_1=0}

x1 / n
y1 / n
t0 / n
m / n
t1 / n

t0 m t1 G:width=30 \gtxt{squ}$_A$
t0 G {-} x1 G {-} ctrl
x1 G:width=30 $+3x^2$ ctrl
t0 m t1 G:width=30 \gtxt{squ}$_A^{-1}$

%% file: quantum-circuits/Qpic/kaliski
PREAMBLE \providecommand{\ket}[1]{\left\vert #1\right\rangle}
PREAMBLE \providecommand{\gtxt}[1]{\texttt{#1}}

carry W \ket{\gtxt{carry}=0} \ket{0}
u W \ket{u_i} \ket{u_{i+1}}
v W \ket{v_i} \ket{v_{i+1}}
r W \ket{r_i} \ket{r_{i+1}}
s W \ket{s_i} \ket{r_{i+1}}
b W \ket{\gtxt{b}=0} \ket{0}
a W \ket{\gtxt{a}=0} \ket{0}
m W \ket{m_i=0} \ket{m_i}

u / n
v / n
r / n+1
s / n+1

r G . fill=none color=white height=0 width=0
u TOUCH

-u +a
-v -a +m
a +b
m +b
carry u v G {-}
carry -b +a +m
m +b
a +b
carry u v G {+}
u G:width=20 $/2$ s G:width=20 $\times 2$ a -m
v G:width=20 $/2$ r G:width=20 $\times 2$ -a m

u v G {-} r s G {+} a m
u G:width=20 $/2$ s G:width=20 $\times 2$ a m

u v G {-} r s G {+} a m
v G:width=20 $/2$ r G:width=20 $\times 2$ -a -m

r +a

%% file: quantum-circuits/Qpic/kaliski-efficient
PREAMBLE \providecommand{\ket}[1]{\left\vert #1\right\rangle}
PREAMBLE \providecommand{\gtxt}[1]{\texttt{#1}}

carry W \ket{\gtxt{carry}=0} \ket{0}
u W \ket{u_i} \ket{u_{i+1}}
v W \ket{v_i} \ket{v_{i+1}}
r W \ket{r_i} \ket{r_{i+1}}
s W \ket{s_i} \ket{r_{i+1}}
b W \ket{\gtxt{b}=0} \ket{0}
a W \ket{\gtxt{a}=0} \ket{0}
m W \ket{m_i=0} \ket{m_i}

u / n
v / n
r / n+1
s / n+1

r G . fill=none color=white height=0 width=0
u TOUCH

-u +a
-v -a +m
u v G $>$ +carry
a +b
m +b
carry -b +a +m
m +b
a +b
u v G $>$ +carry
u v SWAP a
r s SWAP a

m +a

u v G {$-$} r s G {+} a
v TOUCH
m +a

v G:width=20 $/2$
r G:width=20 $\times 2$

r s SWAP a
u v SWAP a

r +a

%% file: quantum-circuits/Qpic/ECC-div
PREAMBLE \providecommand{\ket}[1]{\left\vert #1\right\rangle}
PREAMBLE \providecommand{\gtxt}[1]{\texttt{#1}}

x1 W \ket{x_1}
y1 W \ket{y_1}
ctrl W \ket{\gtxt{ctrl}}
t0 W \ket{t_0=0}
m W \ket{\lambda}

x1 / n
y1 / n
t0 / n
m / n

x1 G \gtxt{inv} width=20 t0 G \gtxt{inv} width=20 
y1 G \gtxt{mul} width=20 t0 G \gtxt{mul} width=20 m G \gtxt{mul} width=20 
x1 G \gtxt{mul} width=20 y1 G \gtxt{mul} width=20 m G \gtxt{mul} width=20 
x1 G \gtxt{inv} width=20 t0 G \gtxt{inv} width=20 

%% file: quantum-circuits/Qpic/ECC-div-expanded
PREAMBLE \providecommand{\ket}[1]{\left\vert #1\right\rangle}
PREAMBLE \providecommand{\gtxt}[1]{\texttt{#1}}

x1 W \ket{x_1}
y1 W \ket{y_1}
ctrl W \ket{\gtxt{ctrl}}
m W \ket{\lambda=0}
t0 W
t1 W
t2 W
m1 W
m2 W

x1 / n
y1 / n
m / n

t0 IN 0
t1 IN 0
t2 IN 0
m1 IN 0
m2 IN 0

x1 G \gtxt{inv}$_A$ width=30 height=20 t0 t1 t2 m1 m2 G \gtxt{inv}$_A$ 

t0 OUT 0
t1 OUT 0
t2 OUT 0
t0 IN 0
t1 IN 0

x1 y1 G \gtxt{mul}$_A$ width=30 t0 t1 G \gtxt{mul}$_A$ width=30
ctrl +m t0
x1 y1 G \gtxt{mul}$_A^{-1}$ width=30 height=15 t0 t1 G \gtxt{mul}$_A^{-1}$ width=30

t0 OUT 0
t1 OUT 0
t2 TOUCH

t0 IN 0
t1 IN 0
t2 IN 0

x1 G \gtxt{inv}$_A^{-1}$ width=30 height=20 t0 t1 t2 m1 m2 G \gtxt{inv}$_A^{-1}$ 

t0 OUT 0
t1 OUT 0
t2 OUT 0
m1 OUT 0
m2 OUT 0

%% file: elliptic-curves.tex
% !TeX root = ./ECDLP.tex
Elliptic curve arithmetic has been heavily optimized for \emph{classical} computers, but because Shor's algorithm requires unique representations of points and in-place point addition, few of these optimizations apply. We find affine Weierstrass coordinates to be the most efficient method; Appendix~\ref{app:curve_representations} explains this choice in more detail.

Affine Weierstrass coordinates are one of the conceptually simplest methods, and RNSL use them for their circuit. We assume the elliptic curve equation has the form 
$E:y^2 = x^3 + ax + b$ with $a,b \in \F_p$
and we represent points by pairs $(x,y)$ that satisfy this equation. Combining RNSL's circuit with the optimizations from this paper, we obtain \iffullversion{}the circuit in Figure \ref{fig:ECC_full}. The total cost is\else{}a circuit with a total cost of\fi{} 2 divisions, 2 multiplications, 1 squaring, and 9 additions. 

The simple formulas with affine coordinates are naturally in-place. The general concept is the same as in \cite{rnsl2017}. The original $x$ and $y$ coordinates are replaced by multiplying and adding to them, and once we have, we can use the new coordinates to uncompute the slope. This also means that the circuit produces incorrect outputs if $P$ or $Q$ is the point at infinity or if $P=\pm Q$,
because in these cases the slope does not exist. Proos and Zalka \cite{pz2003} and RNSL \cite{rnsl2017} both argue that these exceptional cases only slightly distort the desired quantum state in Shor's algorithm and their influence is negligible.

\iffullversion
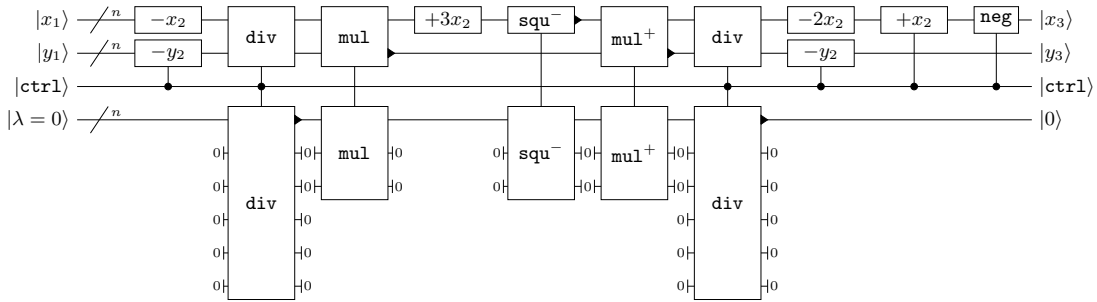
\begin{figure}
\resizebox{\textwidth}{!}{
\input{quantum-circuits/Qpic/ECC-full}
}
\caption{A single elliptic curve addition of a classical point $P=(x_2,y_2)$ with a quantum point $Q=(x_1,y_1)$ in affine Weierstrass coordinates, with $P+Q=(x_3,y_3)$. $\gtxt{div}$ indicates the circuit from Figure \ref{fig:div-efficient}, $\gtxt{squ}^-$ indicates squaring-and-subtracting (as in Figure \ref{fig:square_efficient}) and $\gtxt{mul}^+$ indicates multiplying-and-adding.}\label{fig:ECC_full}
\end{figure}
\fi

\subsection{Windowed arithmetic}\label{sec:point_addition_windowing}
Shor's algorithm uses $2n$ qubits as control qubits, half of which control circuits that add $P$, $2P$, $4P$, etc.; the other half control addition of $Q$, $2Q$, etc. The points are added to a single quantum register which we call the \emph{accumulator}. This process requires $2n$ point additions. We instead use a windowed approach, analogous to ubiquitous classical pre-computation techniques and similar to the techniques used for RSA in \cite{ge2019}. Other classical pre-computation strategies are less effective, as Appendix~\ref{app:precomputation} discusses.

For windowed scalar multiplication, we use the index as an address for a sequential quantum look-up, which loads a superposition of points $\style{O}, P,2P,3P,\dots,(2^\ell - 1)P$ into an auxiliary register which we call the \emph{cache}. For elliptic curves, this requires us to switch from a circuit adding a classical point to a circuit adding two quantum points, but this has little effect on the cost. The depth and $T$-count of the look-up are exponential in the window size $\ell$, but we save $\ell$ point additions. 

\subsubsection{Signed addition.}
We can save a factor of 2 in windowing by using one qubit to control the sign of $P=(x,y)$ and only using $\ell-1$ for the look-up. Since $-P=(x,-y)$, the extra qubit only needs to control a cheap modular negation. This changes the indexing slightly. Our original windowed circuit took an address register $\ket{b}$ and an input register $\ket{R}$, and produced $\ket{b}\ket{[b]P+R}$. Using the top bit $b_{\ell-1}$ of $b=b_{\ell-1}2^{\ell-1} + b'$ as a sign and looking up $b'P$ if $b_{\ell-1} = 1$ and $(2^{\ell-1} - b')P$ and negating it if $b_{\ell-1} = 0$, we can implement the operation
\begin{equation}
\sum_{b\in \{0,1\}^\ell} \ket{b}\ket{R} \mapsto \sum_{b\in\{0,1\}^\ell}\ket{b}\ket{[b-2^{\ell-1}]P+R}.
\end{equation}
Thus, we get an offset in each round, but since the offset is constant, it has no effect on the final phase estimation.

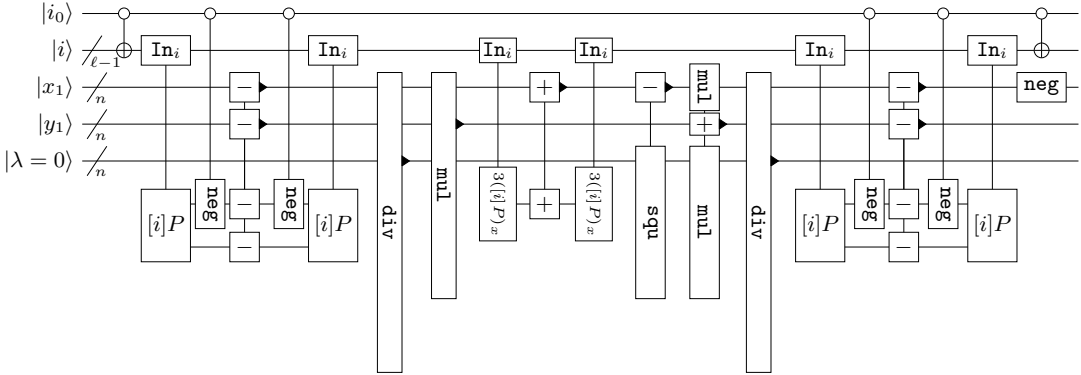
\begin{figure}
\resizebox{\textwidth}{!}{
\input{quantum-circuits/Qpic/ECC-signed-windowed-mod}
}
\caption{Signed windowed elliptic curve addition, with $\ell$ address qubits. $\mathsf{In}_i$ is the indexing half of the look-up gate, which writes the classical data shown into the output registers. The width of the circuits is proportional to the number of auxiliary qubits used.}\label{fig:ECC-windowed}
\end{figure}

%%% Local Variables:
%%% mode: latex
%%% TeX-master: "ECDLP"
%%% End:

%% file: quantum-circuits/Qpic/ECC-full
PREAMBLE \providecommand{\ket}[1]{\left\vert #1\right\rangle}
PREAMBLE \providecommand{\gtxt}[1]{\texttt{#1}}

x1 W \ket{x_1}
y1 W \ket{y_1}
ctrl W \ket{\gtxt{ctrl}}
m W \ket{\lambda=0}
t0 W
t1 W
t2 W
t3 W
t4 W

x1 / n
y1 / n
m / n

x1 G $-x_2$ width=30
y1 G $-y_2$ width=30 ctrl
x1 y1 G \gtxt{div} m t0 t1 t2 t3 t4 G \gtxt{div} ctrl width=30
x1 y1 G \gtxt{mul} m t0 t1 G \gtxt{mul} width=30
x1 G $+3x_2$ width=30
x1 G \gtxt{squ}$^{-}$ m t0 t1 G \gtxt{squ}$^{-}$ width=30
x1 y1 G \gtxt{mul}$^+$ m t0 t1 G \gtxt{mul}$^+$ width=30
x1 y1 G \gtxt{div} m t0 t1 t2 t3 t4 G \gtxt{div} ctrl width=30
x1 G $-2x_2$ width=30
y1 G $-y_2$ width=30 ctrl
x1 G $+x_2$ width=30 ctrl
x1 G \gtxt{neg} width=20 ctrl

%% file: quantum-circuits/Qpic/ECC-signed-windowed-mod.tex
\begin{tikzpicture}[scale=1.000000,x=1pt,y=1pt]
\filldraw[color=white] (0.000000, -7.500000) rectangle (412.000000, 162.500000);
% Drawing wires
% Line 6: sign W \ket{i_0}
\draw[color=black] (0.000000,155.000000) -- (405.000000,155.000000);
\draw[color=black] (0.000000,155.000000) node[left] {$\ket{i_0}$};
% Line 7: address W \ket{i}
\draw[color=black] (0.000000,140.000000) -- (405.000000,140.000000);
\draw[color=black] (0.000000,140.000000) node[left] {$\ket{i}$};
% Line 8: x1 W \ket{x_1}
\draw[color=black] (0.000000,125.000000) -- (405.000000,125.000000);
\draw[color=black] (0.000000,125.000000) node[left] {$\ket{x_1}$};
% Line 9: y1 W \ket{y_1}
\draw[color=black] (0.000000,110.000000) -- (405.000000,110.000000);
\draw[color=black] (0.000000,110.000000) node[left] {$\ket{y_1}$};
% Line 10: m W \ket{\lambda=0}
\draw[color=black] (0.000000,95.000000) -- (405.000000,95.000000);
\draw[color=black] (0.000000,95.000000) node[left] {$\ket{\lambda=0}$};

\ancilla{24}{111}{77.5}
\ancilla{24}{111}{60}

\ancilla{121}{131}{77.5}
\ancilla{121}{131}{60}
\ancilla{121}{131}{45}
\ancilla{121}{131}{30}
\ancilla{121}{131}{15}

\ancilla{141}{151.75}{77.5}
\ancilla{141}{151.75}{60}
\ancilla{141}{151.75}{45}

\ancilla{161.75}{215.25}{77.5}

\ancilla{225.25}{237}{77.5}
\ancilla{225.25}{237}{60}
\ancilla{225.25}{237}{45}

\ancilla{247}{259.5}{77.5}
\ancilla{247}{259.5}{60}
\ancilla{247}{259.5}{45}

\ancilla{269.5}{280}{77.5}
\ancilla{269.5}{280}{60}
\ancilla{269.5}{280}{45}
\ancilla{269.5}{280}{30}
\ancilla{269.5}{280}{15}

\ancilla{290}{380}{77.5}
\ancilla{290}{380}{60}

%% Line 11: t0 W breadth=20
%\draw[color=black] (0.000000,77.500000) -- (412.000000,77.500000);
%% Line 12: t1 W
%\draw[color=black] (0.000000,60.000000) -- (412.000000,60.000000);
%% Line 13: t2 W
%\draw[color=black] (0.000000,45.000000) -- (412.000000,45.000000);
%% Line 14: t3 W
%\draw[color=black] (0.000000,30.000000) -- (412.000000,30.000000);
%% Line 15: t4 W
%\draw[color=black] (0.000000,15.000000) -- (412.000000,15.000000);
%% Line 17: bonus W
%\draw[color=black] (0.000000,0.000000) -- (412.000000,0.000000);
% Done with wires; drawing gates
% Line 19: address / \ell
\draw (0.000000, 134.000000) -- (8.000000, 146.000000);
\draw (0.000000, 135.000000) node[right] {$\scriptstyle{\ell-1}$};
% Line 20: x1 / n
\draw (2.000000, 119.000000) -- (10.000000, 131.000000);
\draw (2.000000, 120.000000) node[right] {$\scriptstyle{n}$};
% Line 21: y1 / n
\draw (2.000000, 104.000000) -- (10.000000, 116.000000);
\draw (2.000000, 105.000000) node[right] {$\scriptstyle{n}$};
% Line 22: m / n
\draw (2.000000, 89.000000) -- (10.000000, 101.000000);
\draw (2.000000, 90.000000) node[right] {$\scriptstyle{n}$};
% Line 24: -sign +address
\draw (17.000000,155.000000) -- (17.000000,140.000000);
\draw[fill=white] (17.000000, 155.000000) circle(2.250000pt);
\begin{scope}
\draw[fill=white] (17.000000, 140.000000) circle(3.000000pt);
\clip (17.000000, 140.000000) circle(3.000000pt);
\draw (14.000000, 140.000000) -- (20.000000, 140.000000);
\draw (17.000000, 137.000000) -- (17.000000, 143.000000);
\end{scope}
% Line 26: address G $\gtxt{In}_i$ t0 t1 G $[i]P$ width=20
\draw (34.000000,140.000000) -- (34.000000,60.000000);
\begin{scope}
\draw[fill=white] (34.000000, 140.000000) +(-45.000000:14.142136pt and 8.485281pt) -- +(45.000000:14.142136pt and 8.485281pt) -- +(135.000000:14.142136pt and 8.485281pt) -- +(225.000000:14.142136pt and 8.485281pt) -- cycle;
\clip (34.000000, 140.000000) +(-45.000000:14.142136pt and 8.485281pt) -- +(45.000000:14.142136pt and 8.485281pt) -- +(135.000000:14.142136pt and 8.485281pt) -- +(225.000000:14.142136pt and 8.485281pt) -- cycle;
\draw (34.000000, 140.000000) node {$\gtxt{In}_i$};
\end{scope}
\begin{scope}
\draw[fill=white] (34.000000, 68.750000) +(-45.000000:14.142136pt and 20.859650pt) -- +(45.000000:14.142136pt and 20.859650pt) -- +(135.000000:14.142136pt and 20.859650pt) -- +(225.000000:14.142136pt and 20.859650pt) -- cycle;
\clip (34.000000, 68.750000) +(-45.000000:14.142136pt and 20.859650pt) -- +(45.000000:14.142136pt and 20.859650pt) -- +(135.000000:14.142136pt and 20.859650pt) -- +(225.000000:14.142136pt and 20.859650pt) -- cycle;
\draw (34.000000, 68.750000) node {$[i]P$};
\end{scope}
% Line 28: t0 G \rotatebox{270}{\gtxt{neg}} width=12 breadth=20 -sign
\draw (52.000000,155.000000) -- (52.000000,77.500000);
\begin{scope}
\draw[fill=white] (52.000000, 77.500000) +(-45.000000:8.485281pt and 14.142136pt) -- +(45.000000:8.485281pt and 14.142136pt) -- +(135.000000:8.485281pt and 14.142136pt) -- +(225.000000:8.485281pt and 14.142136pt) -- cycle;
\clip (52.000000, 77.500000) +(-45.000000:8.485281pt and 14.142136pt) -- +(45.000000:8.485281pt and 14.142136pt) -- +(135.000000:8.485281pt and 14.142136pt) -- +(225.000000:8.485281pt and 14.142136pt) -- cycle;
\draw (52.000000, 77.500000) node {\rotatebox{270}{\gtxt{neg}}};
\end{scope}
\draw[fill=white] (52.000000, 155.000000) circle(2.250000pt);
% Line 28: t0 G \rotatebox{270}{\gtxt{neg}} width=12 breadth=20 -sign
\draw (84.000000,155.000000) -- (84.000000,77.500000);
\begin{scope}
\draw[fill=white] (84.000000, 77.500000) +(-45.000000:8.485281pt and 14.142136pt) -- +(45.000000:8.485281pt and 14.142136pt) -- +(135.000000:8.485281pt and 14.142136pt) -- +(225.000000:8.485281pt and 14.142136pt) -- cycle;
\clip (84.000000, 77.500000) +(-45.000000:8.485281pt and 14.142136pt) -- +(45.000000:8.485281pt and 14.142136pt) -- +(135.000000:8.485281pt and 14.142136pt) -- +(225.000000:8.485281pt and 14.142136pt) -- cycle;
\draw (84.000000, 77.500000) node {\rotatebox{270}{\gtxt{neg}}};
\end{scope}
\draw[fill=white] (84.000000, 155.000000) circle(2.250000pt);
% Line 30: y1 G $-$ t1 G $-$
\draw (66.000000,125.000000) -- (66.000000,60.000000);
\begin{scope}
\draw[fill=white] (66.000000, 125.000000) +(-45.000000:8.485281pt and 8.485281pt) -- +(45.000000:8.485281pt and 8.485281pt) -- +(135.000000:8.485281pt and 8.485281pt) -- +(225.000000:8.485281pt and 8.485281pt) -- cycle;
\clip (66.000000, 125.000000) +(-45.000000:8.485281pt and 8.485281pt) -- +(45.000000:8.485281pt and 8.485281pt) -- +(135.000000:8.485281pt and 8.485281pt) -- +(225.000000:8.485281pt and 8.485281pt) -- cycle;
\draw (66.000000, 125.000000) node {$-$};
\end{scope}
\begin{scope}
\draw[fill=white] (66.000000, 60.000000) +(-45.000000:8.485281pt and 8.485281pt) -- +(45.000000:8.485281pt and 8.485281pt) -- +(135.000000:8.485281pt and 8.485281pt) -- +(225.000000:8.485281pt and 8.485281pt) -- cycle;
\clip (66.000000, 60.000000) +(-45.000000:8.485281pt and 8.485281pt) -- +(45.000000:8.485281pt and 8.485281pt) -- +(135.000000:8.485281pt and 8.485281pt) -- +(225.000000:8.485281pt and 8.485281pt) -- cycle;
\draw (66.000000, 60.000000) node {$-$};
\end{scope}
% Line 29: x1 G $-$ t0 G $-$
\draw (66.000000,110.000000) -- (66.000000,77.500000);
\begin{scope}
\draw[fill=white] (66.000000, 110.000000) +(-45.000000:8.485281pt and 8.485281pt) -- +(45.000000:8.485281pt and 8.485281pt) -- +(135.000000:8.485281pt and 8.485281pt) -- +(225.000000:8.485281pt and 8.485281pt) -- cycle;
\clip (66.000000, 110.000000) +(-45.000000:8.485281pt and 8.485281pt) -- +(45.000000:8.485281pt and 8.485281pt) -- +(135.000000:8.485281pt and 8.485281pt) -- +(225.000000:8.485281pt and 8.485281pt) -- cycle;
\draw (66.000000, 110.000000) node {$-$};
\end{scope}
\begin{scope}
\draw[fill=white] (66.000000, 77.500000) +(-45.000000:8.485281pt and 8.485281pt) -- +(45.000000:8.485281pt and 8.485281pt) -- +(135.000000:8.485281pt and 8.485281pt) -- +(225.000000:8.485281pt and 8.485281pt) -- cycle;
\clip (66.000000, 77.500000) +(-45.000000:8.485281pt and 8.485281pt) -- +(45.000000:8.485281pt and 8.485281pt) -- +(135.000000:8.485281pt and 8.485281pt) -- +(225.000000:8.485281pt and 8.485281pt) -- cycle;
\draw (66.000000, 77.500000) node {$-$};
\end{scope}
% Line 32: address G $\gtxt{In}_i$ t0 t1 G $[i]P$ width=20
\draw (102.000000,140.000000) -- (102.000000,60.000000);
\begin{scope}
\draw[fill=white] (102.000000, 140.000000) +(-45.000000:14.142136pt and 8.485281pt) -- +(45.000000:14.142136pt and 8.485281pt) -- +(135.000000:14.142136pt and 8.485281pt) -- +(225.000000:14.142136pt and 8.485281pt) -- cycle;
\clip (102.000000, 140.000000) +(-45.000000:14.142136pt and 8.485281pt) -- +(45.000000:14.142136pt and 8.485281pt) -- +(135.000000:14.142136pt and 8.485281pt) -- +(225.000000:14.142136pt and 8.485281pt) -- cycle;
\draw (102.000000, 140.000000) node {$\gtxt{In}_i$};
\end{scope}
\begin{scope}
\draw[fill=white] (102.000000, 68.750000) +(-45.000000:14.142136pt and 20.859650pt) -- +(45.000000:14.142136pt and 20.859650pt) -- +(135.000000:14.142136pt and 20.859650pt) -- +(225.000000:14.142136pt and 20.859650pt) -- cycle;
\clip (102.000000, 68.750000) +(-45.000000:14.142136pt and 20.859650pt) -- +(45.000000:14.142136pt and 20.859650pt) -- +(135.000000:14.142136pt and 20.859650pt) -- +(225.000000:14.142136pt and 20.859650pt) -- cycle;
\draw (102.000000, 68.750000) node {$[i]P$};
\end{scope}
% Line 33: TOUCH
% Line 34: x1 y1  m t0 t1 t2 t3 t4 G \rotatebox{270}{\gtxt{div}} width=10
\draw (125.000000,125.000000) -- (125.000000,15.000000);
\begin{scope}
\draw[fill=white] (125.000000, 70.000000) +(-45.000000:7.071068pt and 86.267027pt) -- +(45.000000:7.071068pt and 86.267027pt) -- +(135.000000:7.071068pt and 86.267027pt) -- +(225.000000:7.071068pt and 86.267027pt) -- cycle;
\clip (125.000000, 70.000000) +(-45.000000:7.071068pt and 86.267027pt) -- +(45.000000:7.071068pt and 86.267027pt) -- +(135.000000:7.071068pt and 86.267027pt) -- +(225.000000:7.071068pt and 86.267027pt) -- cycle;
\draw (125.000000, 70.000000) node {\rotatebox{270}{\gtxt{div}}};
\end{scope}
% Line 37: x1 y1  m t0 t1 t2 G \rotatebox{270}{\gtxt{mul}}	 width=10
\draw (147.000000,125.000000) -- (147.000000,45.000000);
\begin{scope}
\draw[fill=white] (147.000000, 85.000000) +(-45.000000:7.071068pt and 65.053824pt) -- +(45.000000:7.071068pt and 65.053824pt) -- +(135.000000:7.071068pt and 65.053824pt) -- +(225.000000:7.071068pt and 65.053824pt) -- cycle;
\clip (147.000000, 85.000000) +(-45.000000:7.071068pt and 65.053824pt) -- +(45.000000:7.071068pt and 65.053824pt) -- +(135.000000:7.071068pt and 65.053824pt) -- +(225.000000:7.071068pt and 65.053824pt) -- cycle;
\draw (147.000000, 85.000000) node {\rotatebox{270}{\gtxt{mul}}};
\end{scope}
% Line 38: bonus G ? width=18
% Line 42: address G:breadth=10 $\gtxt{In}_i$ t0 G:breadth=30 \rotatebox{270}{$\scriptstyle{3([i]P)_x}$} width=15
\draw (169.000000,140.000000) -- (169.000000,77.500000);
\begin{scope}
\draw[fill=white] (169.000000, 140.000000) +(-45.000000:10.606602pt and 7.071068pt) -- +(45.000000:10.606602pt and 7.071068pt) -- +(135.000000:10.606602pt and 7.071068pt) -- +(225.000000:10.606602pt and 7.071068pt) -- cycle;
\clip (169.000000, 140.000000) +(-45.000000:10.606602pt and 7.071068pt) -- +(45.000000:10.606602pt and 7.071068pt) -- +(135.000000:10.606602pt and 7.071068pt) -- +(225.000000:10.606602pt and 7.071068pt) -- cycle;
\draw (169.000000, 140.000000) node {$\gtxt{In}_i$};
\end{scope}
\begin{scope}
\draw[fill=white] (169.000000, 77.500000) +(-45.000000:10.606602pt and 21.213203pt) -- +(45.000000:10.606602pt and 21.213203pt) -- +(135.000000:10.606602pt and 21.213203pt) -- +(225.000000:10.606602pt and 21.213203pt) -- cycle;
\clip (169.000000, 77.500000) +(-45.000000:10.606602pt and 21.213203pt) -- +(45.000000:10.606602pt and 21.213203pt) -- +(135.000000:10.606602pt and 21.213203pt) -- +(225.000000:10.606602pt and 21.213203pt) -- cycle;
\draw (169.000000, 77.500000) node {\rotatebox{270}{$\scriptstyle{3([i]P)_x}$}};
\end{scope}
% Line 43: x1 G $+$ t0 G $+$
\draw (188.000000,125.000000) -- (188.000000,77.500000);
\begin{scope}
\draw[fill=white] (188.000000, 125.000000) +(-45.000000:8.485281pt and 8.485281pt) -- +(45.000000:8.485281pt and 8.485281pt) -- +(135.000000:8.485281pt and 8.485281pt) -- +(225.000000:8.485281pt and 8.485281pt) -- cycle;
\clip (188.000000, 125.000000) +(-45.000000:8.485281pt and 8.485281pt) -- +(45.000000:8.485281pt and 8.485281pt) -- +(135.000000:8.485281pt and 8.485281pt) -- +(225.000000:8.485281pt and 8.485281pt) -- cycle;
\draw (188.000000, 125.000000) node {$+$};
\end{scope}
\begin{scope}
\draw[fill=white] (188.000000, 77.500000) +(-45.000000:8.485281pt and 8.485281pt) -- +(45.000000:8.485281pt and 8.485281pt) -- +(135.000000:8.485281pt and 8.485281pt) -- +(225.000000:8.485281pt and 8.485281pt) -- cycle;
\clip (188.000000, 77.500000) +(-45.000000:8.485281pt and 8.485281pt) -- +(45.000000:8.485281pt and 8.485281pt) -- +(135.000000:8.485281pt and 8.485281pt) -- +(225.000000:8.485281pt and 8.485281pt) -- cycle;
\draw (188.000000, 77.500000) node {$+$};
\end{scope}

% Line 46: address G:breadth=10  $\gtxt{In}_i$ t0 G:breadth=30 \rotatebox{270}{$\scriptstyle{3([i]P)_x}$} width=15
\draw (208.000000,140.000000) -- (208.000000,77.500000);
\begin{scope}
\draw[fill=white] (208.000000, 140.000000) +(-45.000000:10.606602pt and 7.071068pt) -- +(45.000000:10.606602pt and 7.071068pt) -- +(135.000000:10.606602pt and 7.071068pt) -- +(225.000000:10.606602pt and 7.071068pt) -- cycle;
\clip (208.000000, 140.000000) +(-45.000000:10.606602pt and 7.071068pt) -- +(45.000000:10.606602pt and 7.071068pt) -- +(135.000000:10.606602pt and 7.071068pt) -- +(225.000000:10.606602pt and 7.071068pt) -- cycle;
\draw (208.000000, 140.000000) node {$\gtxt{In}_i$};
\end{scope}
\begin{scope}
\draw[fill=white] (208.000000, 77.500000) +(-45.000000:10.606602pt and 21.213203pt) -- +(45.000000:10.606602pt and 21.213203pt) -- +(135.000000:10.606602pt and 21.213203pt) -- +(225.000000:10.606602pt and 21.213203pt) -- cycle;
\clip (208.000000, 77.500000) +(-45.000000:10.606602pt and 21.213203pt) -- +(45.000000:10.606602pt and 21.213203pt) -- +(135.000000:10.606602pt and 21.213203pt) -- +(225.000000:10.606602pt and 21.213203pt) -- cycle;
\draw (208.000000, 77.500000) node {\rotatebox{270}{$\scriptstyle{3([i]P)_x}$}};
\end{scope}
% Line 49: x1 G $-$ m t0 t1 t2 G \rotatebox{270}{\gtxt{squ}}
\draw (231.000000,125.000000) -- (231.000000,45.000000);
\begin{scope}
\draw[fill=white] (231.000000, 125.000000) +(-45.000000:8.485281pt and 8.485281pt) -- +(45.000000:8.485281pt and 8.485281pt) -- +(135.000000:8.485281pt and 8.485281pt) -- +(225.000000:8.485281pt and 8.485281pt) -- cycle;
\clip (231.000000, 125.000000) +(-45.000000:8.485281pt and 8.485281pt) -- +(45.000000:8.485281pt and 8.485281pt) -- +(135.000000:8.485281pt and 8.485281pt) -- +(225.000000:8.485281pt and 8.485281pt) -- cycle;
\draw (231.000000, 125.000000) node {$-$};
\end{scope}
\begin{scope}
\draw[fill=white] (231.000000, 70.000000) +(-45.000000:8.485281pt and 43.840620pt) -- +(45.000000:8.485281pt and 43.840620pt) -- +(135.000000:8.485281pt and 43.840620pt) -- +(225.000000:8.485281pt and 43.840620pt) -- cycle;
\clip (231.000000, 70.000000) +(-45.000000:8.485281pt and 43.840620pt) -- +(45.000000:8.485281pt and 43.840620pt) -- +(135.000000:8.485281pt and 43.840620pt) -- +(225.000000:8.485281pt and 43.840620pt) -- cycle;
\draw (231.000000, 70.000000) node {\rotatebox{270}{\gtxt{squ}}};
\end{scope}
% Line 52: x1 G:breadth=19 \rotatebox{270}{\gtxt{mul}} y1 G:breadth=9 $+$ m t0 t1 t2 G \rotatebox{270}{\gtxt{mul}}
\draw (253.000000,125.000000) -- (253.000000,45.000000);
\begin{scope}
\draw[fill=white] (253.000000, 125.000000) +(-45.000000:8.485281pt and 13.435029pt) -- +(45.000000:8.485281pt and 13.435029pt) -- +(135.000000:8.485281pt and 13.435029pt) -- +(225.000000:8.485281pt and 13.435029pt) -- cycle;
\clip (253.000000, 125.000000) +(-45.000000:8.485281pt and 13.435029pt) -- +(45.000000:8.485281pt and 13.435029pt) -- +(135.000000:8.485281pt and 13.435029pt) -- +(225.000000:8.485281pt and 13.435029pt) -- cycle;
\draw (253.000000, 125.000000) node {\rotatebox{270}{\gtxt{mul}}};
\end{scope}
\begin{scope}
\draw[fill=white] (253.000000, 110.000000) +(-45.000000:8.485281pt and 6.363961pt) -- +(45.000000:8.485281pt and 6.363961pt) -- +(135.000000:8.485281pt and 6.363961pt) -- +(225.000000:8.485281pt and 6.363961pt) -- cycle;
\clip (253.000000, 110.000000) +(-45.000000:8.485281pt and 6.363961pt) -- +(45.000000:8.485281pt and 6.363961pt) -- +(135.000000:8.485281pt and 6.363961pt) -- +(225.000000:8.485281pt and 6.363961pt) -- cycle;
\draw (253.000000, 110.000000) node {$+$};
\end{scope}
\begin{scope}
\draw[fill=white] (253.000000, 70.000000) +(-45.000000:8.485281pt and 43.840620pt) -- +(45.000000:8.485281pt and 43.840620pt) -- +(135.000000:8.485281pt and 43.840620pt) -- +(225.000000:8.485281pt and 43.840620pt) -- cycle;
\clip (253.000000, 70.000000) +(-45.000000:8.485281pt and 43.840620pt) -- +(45.000000:8.485281pt and 43.840620pt) -- +(135.000000:8.485281pt and 43.840620pt) -- +(225.000000:8.485281pt and 43.840620pt) -- cycle;
\draw (253.000000, 70.000000) node {\rotatebox{270}{\gtxt{mul}}};
\end{scope}
% Line 55: x1 y1 m t0 t1 t2 t3 t4 G \rotatebox{270}{\gtxt{div}} width=10
\draw (275.000000,125.000000) -- (275.000000,15.000000);
\begin{scope}
\draw[fill=white] (275.000000, 70.000000) +(-45.000000:7.071068pt and 86.267027pt) -- +(45.000000:7.071068pt and 86.267027pt) -- +(135.000000:7.071068pt and 86.267027pt) -- +(225.000000:7.071068pt and 86.267027pt) -- cycle;
\clip (275.000000, 70.000000) +(-45.000000:7.071068pt and 86.267027pt) -- +(45.000000:7.071068pt and 86.267027pt) -- +(135.000000:7.071068pt and 86.267027pt) -- +(225.000000:7.071068pt and 86.267027pt) -- cycle;
\draw (275.000000, 70.000000) node {\rotatebox{270}{\gtxt{div}}};
\end{scope}
% Line 58: address G $\gtxt{In}_i$ t0 t1 G $[i]P$ width=20
\draw (300.000000,140.000000) -- (300.000000,60.000000);
\begin{scope}
\draw[fill=white] (300.000000, 140.000000) +(-45.000000:14.142136pt and 8.485281pt) -- +(45.000000:14.142136pt and 8.485281pt) -- +(135.000000:14.142136pt and 8.485281pt) -- +(225.000000:14.142136pt and 8.485281pt) -- cycle;
\clip (300.000000, 140.000000) +(-45.000000:14.142136pt and 8.485281pt) -- +(45.000000:14.142136pt and 8.485281pt) -- +(135.000000:14.142136pt and 8.485281pt) -- +(225.000000:14.142136pt and 8.485281pt) -- cycle;
\draw (300.000000, 140.000000) node {$\gtxt{In}_i$};
\end{scope}
\begin{scope}
\draw[fill=white] (300.000000, 68.750000) +(-45.000000:14.142136pt and 20.859650pt) -- +(45.000000:14.142136pt and 20.859650pt) -- +(135.000000:14.142136pt and 20.859650pt) -- +(225.000000:14.142136pt and 20.859650pt) -- cycle;
\clip (300.000000, 68.750000) +(-45.000000:14.142136pt and 20.859650pt) -- +(45.000000:14.142136pt and 20.859650pt) -- +(135.000000:14.142136pt and 20.859650pt) -- +(225.000000:14.142136pt and 20.859650pt) -- cycle;
\draw (300.000000, 68.750000) node {$[i]P$};
\end{scope}
% Line 64: t0 G \rotatebox{270}{\gtxt{neg}} -sign width=12 breadth=20
\draw (320.000000,155.000000) -- (320.000000,77.500000);
\begin{scope}
\draw[fill=white] (320.000000, 77.500000) +(-45.000000:8.485281pt and 14.142136pt) -- +(45.000000:8.485281pt and 14.142136pt) -- +(135.000000:8.485281pt and 14.142136pt) -- +(225.000000:8.485281pt and 14.142136pt) -- cycle;
\clip (320.000000, 77.500000) +(-45.000000:8.485281pt and 14.142136pt) -- +(45.000000:8.485281pt and 14.142136pt) -- +(135.000000:8.485281pt and 14.142136pt) -- +(225.000000:8.485281pt and 14.142136pt) -- cycle;
\draw (320.000000, 77.500000) node {\rotatebox{270}{\gtxt{neg}}};
\end{scope}
\draw[fill=white] (320.000000, 155.000000) circle(2.250000pt);

\draw (350.000000,155.000000) -- (350.000000,77.500000);
\begin{scope}
\draw[fill=white] (350.000000, 77.500000) +(-45.000000:8.485281pt and 14.142136pt) -- +(45.000000:8.485281pt and 14.142136pt) -- +(135.000000:8.485281pt and 14.142136pt) -- +(225.000000:8.485281pt and 14.142136pt) -- cycle;
\clip (350.000000, 77.500000) +(-45.000000:8.485281pt and 14.142136pt) -- +(45.000000:8.485281pt and 14.142136pt) -- +(135.000000:8.485281pt and 14.142136pt) -- +(225.000000:8.485281pt and 14.142136pt) -- cycle;
\draw (350.000000, 77.500000) node {\rotatebox{270}{\gtxt{neg}}};
\end{scope}
\draw[fill=white] (350.000000, 155.000000) circle(2.250000pt);

% Line 67: x1 G $-$ t0 G $-$

% Line 66: y1 G $-$ t1 G $-$
\draw (334.000000,125.000000) -- (334.000000,60.000000);
\begin{scope}
\draw[fill=white] (334.000000, 125.000000) +(-45.000000:8.485281pt and 8.485281pt) -- +(45.000000:8.485281pt and 8.485281pt) -- +(135.000000:8.485281pt and 8.485281pt) -- +(225.000000:8.485281pt and 8.485281pt) -- cycle;
\clip (334.000000, 125.000000) +(-45.000000:8.485281pt and 8.485281pt) -- +(45.000000:8.485281pt and 8.485281pt) -- +(135.000000:8.485281pt and 8.485281pt) -- +(225.000000:8.485281pt and 8.485281pt) -- cycle;
\draw (334.000000, 125.000000) node {$-$};
\end{scope}
\begin{scope}
\draw[fill=white] (334.000000, 60.000000) +(-45.000000:8.485281pt and 8.485281pt) -- +(45.000000:8.485281pt and 8.485281pt) -- +(135.000000:8.485281pt and 8.485281pt) -- +(225.000000:8.485281pt and 8.485281pt) -- cycle;
\clip (334.000000, 60.000000) +(-45.000000:8.485281pt and 8.485281pt) -- +(45.000000:8.485281pt and 8.485281pt) -- +(135.000000:8.485281pt and 8.485281pt) -- +(225.000000:8.485281pt and 8.485281pt) -- cycle;
\draw (334.000000, 60.000000) node {$-$};
\end{scope}
\draw (334.000000,110.000000) -- (334.000000,77.500000);
\begin{scope}
\draw[fill=white] (334.000000, 110.000000) +(-45.000000:8.485281pt and 8.485281pt) -- +(45.000000:8.485281pt and 8.485281pt) -- +(135.000000:8.485281pt and 8.485281pt) -- +(225.000000:8.485281pt and 8.485281pt) -- cycle;
\clip (334.000000, 110.000000) +(-45.000000:8.485281pt and 8.485281pt) -- +(45.000000:8.485281pt and 8.485281pt) -- +(135.000000:8.485281pt and 8.485281pt) -- +(225.000000:8.485281pt and 8.485281pt) -- cycle;
\draw (334.000000, 110.000000) node {$-$};
\end{scope}
\begin{scope}
\draw[fill=white] (334.000000, 77.500000) +(-45.000000:8.485281pt and 8.485281pt) -- +(45.000000:8.485281pt and 8.485281pt) -- +(135.000000:8.485281pt and 8.485281pt) -- +(225.000000:8.485281pt and 8.485281pt) -- cycle;
\clip (334.000000, 77.500000) +(-45.000000:8.485281pt and 8.485281pt) -- +(45.000000:8.485281pt and 8.485281pt) -- +(135.000000:8.485281pt and 8.485281pt) -- +(225.000000:8.485281pt and 8.485281pt) -- cycle;
\draw (334.000000, 77.500000) node {$-$};
\end{scope}
% Line 69: address G $\gtxt{In}_i$ t0 t1 G $[i]P$ width=20
\draw (370.000000,140.000000) -- (370.000000,60.000000);
\begin{scope}
\draw[fill=white] (370.000000, 140.000000) +(-45.000000:14.142136pt and 8.485281pt) -- +(45.000000:14.142136pt and 8.485281pt) -- +(135.000000:14.142136pt and 8.485281pt) -- +(225.000000:14.142136pt and 8.485281pt) -- cycle;
\clip (370.000000, 140.000000) +(-45.000000:14.142136pt and 8.485281pt) -- +(45.000000:14.142136pt and 8.485281pt) -- +(135.000000:14.142136pt and 8.485281pt) -- +(225.000000:14.142136pt and 8.485281pt) -- cycle;
\draw (370.000000, 140.000000) node {$\gtxt{In}_i$};
\end{scope}
\begin{scope}
\draw[fill=white] (370.000000, 68.750000) +(-45.000000:14.142136pt and 20.859650pt) -- +(45.000000:14.142136pt and 20.859650pt) -- +(135.000000:14.142136pt and 20.859650pt) -- +(225.000000:14.142136pt and 20.859650pt) -- cycle;
\clip (370.000000, 68.750000) +(-45.000000:14.142136pt and 20.859650pt) -- +(45.000000:14.142136pt and 20.859650pt) -- +(135.000000:14.142136pt and 20.859650pt) -- +(225.000000:14.142136pt and 20.859650pt) -- cycle;
\draw (370.000000, 68.750000) node {$[i]P$};
\end{scope}
% Line 73: x1 G $\gtxt{neg}$ width=20
\begin{scope}
\draw[fill=white] (390.000000, 125.000000) +(-45.000000:14.142136pt and 8.485281pt) -- +(45.000000:14.142136pt and 8.485281pt) -- +(135.000000:14.142136pt and 8.485281pt) -- +(225.000000:14.142136pt and 8.485281pt) -- cycle;
\clip (390.000000, 125.000000) +(-45.000000:14.142136pt and 8.485281pt) -- +(45.000000:14.142136pt and 8.485281pt) -- +(135.000000:14.142136pt and 8.485281pt) -- +(225.000000:14.142136pt and 8.485281pt) -- cycle;
\draw (390.000000, 125.000000) node {$\gtxt{neg}$};
\end{scope}
% Line 71: +address -sign
\draw (390.000000,155.000000) -- (390.000000,140.000000);
\begin{scope}
\draw[fill=white] (390.000000, 140.000000) circle(3.000000pt);
\clip (390.000000, 140.000000) circle(3.000000pt);
\draw (387.000000, 140.000000) -- (393.000000, 140.000000);
\draw (390.000000, 137.000000) -- (390.000000, 143.000000);
\end{scope}
\draw[fill=white] (390.000000, 155.000000) circle(2.250000pt);
% Done with gates; drawing ending labels
% Done with ending labels; drawing cut lines and comments
% Done with comments

\qOutput{72}{125}{black}
\qOutput{72}{110}{black}

\qOutput{130}{95}{black}
\qOutput{152}{110}{black}

\qOutput{194}{125}{black}

\qOutput{237}{125}{black}
\qOutput{259}{110}{black}
\qOutput{280}{95}{black}

\qOutput{340}{125}{black}
\qOutput{340}{110}{black}

\end{tikzpicture}

%% file: numerical-estimates.tex
We present quantum resource estimates for Shor's algorithm based on our cost estimates for windowed elliptic curve point addition. Results are obtained for optimizing three different cost metrics: either minimizing circuit width (i.e. the total number of logical qubits), the total number of $T$ gates in the circuit, or total circuit depth.
Window sizes above 18 were too large to simulate, so we extrapolated from the costs for smaller lookup tables. With {\qsharp}, we calculated the cost of a point addition on the three NIST curves \cite{fips186-4} P256, P384 and P521, using $8$-bit look-ups, then subtracted the cost of six $8$-bit look-ups and added the cost of six $\ell$-bit look-ups to get the cost of point addition with an $\ell$-bit window. We multiplied this cost by the number of windows dividing $2n$ to get the full cost of Shor's algorithm for ECDLP.  From this we manually selected optimal window sizes. We can use $n$ instead of $n+1$ because the order of each NIST curve is less than its modulus \cite{fips186-4}.

Table~\ref{table:curve_results_continued} shows our results together with those of RNSL \cite{rnsl2017} for comparison. Their circuits use fewer than $2$ Toffoli gates per time step on average, so we assume that with 8 extra qubits (c.f. Section~\ref{seq:toffoli}) they can use Toffoli gates of T-depth 1. Our optimization for the number of qubits is shown in the row labeled \emph{Low W}. Since RNSL optimize for the same metric, those results allow a direct comparison. We were able to improve on RNSL's circuit in all metrics. For P256, we reduce the number of logical qubits from 2338 to 2124, while reducing the $T$-depth and $T$-count by factors of 54 and 119, respectively.

Additionally, we report more significant improvements over RNSL's work in depth and $T$-count when optimizing for those. For 521-bit moduli, the improvement is a factor of 13,792 in depth for an increase of 22\% in qubits, or a 463 factor reduction in $T$-gates for a 12\% increase in qubits. 

\begin{table}
\centering
\begin{tabular}{|| c|| c || c | c|c|c||c |c ||c||}
\hline
\multirow{2}{*}{Circuit} & Window  &\multicolumn{4}{| c||}{Gates} & \multicolumn{2}{| c||}{Depth} & Width\\\cline{3-9}
 &  Size & Cliffords & Measure & T & Total &  T & All gates  & Qubits \\
 \hline
 \multicolumn{9}{| c|}{256-bit modulus}\\
\hline	
RNSL & -- &-- &-- &$1.60\cdot 2^{39}$ & -- &$1.69\cdot 2^{36}$ & -- &$2338$\\
Low W & $19$ & $1.32\cdot 2^{34}$ & $1.76\cdot 2^{26}$ & $1.72\cdot 2^{32}$ & $1.45\cdot 2^{35}$ & $1.98\cdot 2^{30}$ & $1.89\cdot 2^{32}$ & $2124$\\
Low T & $19$ & $1.75\cdot 2^{33}$ & $1.95\cdot 2^{27}$ & $1.08\cdot 2^{31}$ & $1.80\cdot 2^{34}$ & $1.44\cdot 2^{29}$ & $1.85\cdot 2^{31}$ & $2619$\\
Low D & $15$ & $1.04\cdot 2^{34}$ & $1.61\cdot 2^{28}$ & $1.34\cdot 2^{32}$ & $1.40\cdot 2^{34}$ & $1.12\cdot 2^{24}$ & $1.40\cdot 2^{27}$ & $2871$\\
\hline
 \hline
 \multicolumn{9}{| c|}{384-bit modulus}\\
\hline	
RNSL & -- &-- &-- &$1.44\cdot 2^{41}$ & -- &$1.51\cdot 2^{38}$ & -- &$3492$\\
Low W & $21$ & $1.46\cdot 2^{36}$ & $1.23\cdot 2^{29}$ & $1.51\cdot 2^{34}$ & $1.57\cdot 2^{37}$ & $1.68\cdot 2^{32}$ & $1.77\cdot 2^{34}$ & $3151$\\
Low T & $19$ & $1.05\cdot 2^{35}$ & $1.28\cdot 2^{29}$ & $1.74\cdot 2^{32}$ & $1.10\cdot 2^{36}$ & $1.21\cdot 2^{31}$ & $1.31\cdot 2^{33}$ & $3901$\\
Low D & $15$ & $1.73\cdot 2^{35}$ & $1.34\cdot 2^{30}$ & $1.13\cdot 2^{34}$ & $1.17\cdot 2^{36}$ & $1.23\cdot 2^{25}$ & $1.48\cdot 2^{28}$ & $4278$\\
\hline
 \hline
 \multicolumn{9}{| c|}{521-bit modulus}\\
\hline	
RNSL & -- &-- &-- &$1.81\cdot 2^{42}$ & -- &$1.91\cdot 2^{39}$ & -- &$4727$\\
Low W & $22$ & $1.85\cdot 2^{37}$ & $1.59\cdot 2^{30}$ & $1.82\cdot 2^{35}$ & $1.98\cdot 2^{38}$ & $1.99\cdot 2^{33}$ & $1.09\cdot 2^{36}$ & $4258$\\
Low T & $20$ & $1.45\cdot 2^{35}$ & $1.49\cdot 2^{30}$ & $1.00\cdot 2^{34}$ & $1.57\cdot 2^{36}$ & $1.40\cdot 2^{32}$ & $1.54\cdot 2^{34}$ & $5273$\\
Low D & $15$ & $1.10\cdot 2^{37}$ & $1.70\cdot 2^{31}$ & $1.43\cdot 2^{35}$ & $1.48\cdot 2^{37}$ & $1.13\cdot 2^{26}$ & $1.27\cdot 2^{29}$ & $5789$\\
\hline
\end{tabular}
\caption{Resource estimates for Shor's full algorithm to compute the ECDLP. RNSL results are taken from \cite{rnsl2017}. Rows labeled Low W/Low T/ Low D show estimates for circuits minimizing width, $T$ gate count and total depth, respectively.}\label{table:curve_results_continued}
\end{table}

\iffullversion
\subsubsection{Asymptotic formulas.} 
We ran resource estimates for the various components of the elliptic curve point addition circuits on a range of input bit sizes. We then determined models for fitting the data with linear regression and deduced asymptotic formulas for the costs of these operations parameterized by the bit size $n$. Tables~\ref{table:arithmetic}, \ref{table:alternate_arithmetic} and \ref{table:modular_arithmetic} in Appendix~\ref{app:asymptotic} show the results of our experiments. 

Table~\ref{table:modular_arithmetic} also shows such formulas for a full signed windowed elliptic curve point addition and the full circuit of Shor's algorithm using such point addition. The results show that our circuit optimized for low width can solve the ECDLP on an $n$-bit elliptic curve with about $8n+10.2\floor{\lg n}-1$ logical qubits using roughly $436n^3-1.05\cdot 2^{26}$ $T$-gates at a $T$-depth of $120n^3-1.67\cdot 2^{22}$. The total number of gates is $2900n^3 -1.08\cdot 2^{31}$ with depth $509n^3-1.84\cdot 2^{27}$.

Our circuits optimized for a low number of $T$ gates require about $10n+7.4\floor{\lg n} +1.3$ qubits and need $1115n^3/\lg n-1.08\cdot 2^{24}$ $T$ gates with $T$ depth $389n^3/\lg n-1.70\cdot 2^{22}$. Optimizing for low depth brings down the overall depth to $2523n^2+1.10\cdot 2^{20}$ and the $T$-depth to $285n^2-1.54\cdot 2^{17}$, but requires $11n+3.9\floor{\lg n}+16.5$ qubits.
\fi

%% file: appendix.tex
% !TeX root = ./ECDLP.tex
\section{Alternative approaches}
\subsection{Modular multiplication.}\label{app:multiplication_variants}
RNSL provide two circuits for modular multiplication. The first is the one proposed by Proos and Zalka \cite{pz2003}, which uses a double-and-add approach, where doubling and addition are both modular operations modulo $p$. The other is reversible Montgomery multiplication, which uses an add-and-halve approach and works in Montgomery form. The primary motivation for considering Montgomery multiplication instead of the straightforward double-and-add method is that modular reduction is achieved by suitable additions to clear lower order bits and divisions by $2$ (i.e. bit rotations) as part of the whole circuit, not delegated to the addition and halving circuits. This results in simpler operations per bit. 

However, Montgomery multiplication has the downside that it entangles with a register of auxiliary qubits which must be cleared. In our case, at every point in an elliptic curve point addition, we have enough spare auxiliary qubits for this. Overall, it is cheaper, even with the Bennett method, and especially with the multiply-then-add technique of Section~\ref{sec:multiplication_pebbling}.

\iffullversion
Rines and Chuang recently provided high-performance modular multipliers \cite{rc2018}. A direct comparison is difficult, since they give costs in Toffoli gates. Our low-T multiplier scales as $38n^2+o(n^2)$ T gates, while their best exact multiplier uses $4n^2$ Toffoli gates, suggesting perhaps $28n^2$ T gates. However, some of these Toffoli gates might be replaceable with AND gates. Their adder seems to use $4.5n$ additions, slightly more than our $2n+O(n/\log n)$ additions. We leave a true comparison and \qsharp implementation for future work.\fi

\subsection{Modular Inversion.}\label{app:inversion_variants}
Proos and Zalka \cite{pz2003} (PZ) gave an approach to modular inversion based on precise control of a bit-shift division operation, with asymptotic complexity of $O(n^2)$. There are $O(n)$ iterations of a \emph{round}. Each round implements conditional logic by computing state qubits, then using those state qubits to control some operations on the integer registers. 

RNSL use a similar round-based construction, which implements a reversible binary extended Euclidean algorithm. As with multiplication, the primary difference between the PZ division and the RNSL division is that PZ's is based on doubling and integer long division, while RNSL's is based on halving and binary operations. The PZ inversion leaves only $O(\lg n)$ auxiliary qubits, while RNSL creates $2n+O(\lg n)$ auxiliary qubits, but PZ has a higher depth and gate cost. 

Naively, the PZ approach uses $5n$ qubits, though they show that, with fidelity loss on the order of $O(n^{-3})$ per round, they require only $2n+8\sqrt{n}+O(\log n)$ qubits. The RNSL approach uses $6n$ qubits. We choose to use the RNSL algorithm. It is exactly correct, so it can be used for higher depth algorithms, and the total T-cost and depth are less than half of the PZ approach.

\subsection{Recursive GCD Algorithms.}
There are several sub-quadratic GCD algorithms (such as \cite{by2019}). These work by defining a series of $2\times 2$ matrices $T_n$ such that $T_nT_{n-1}\cdots T_1(u,v)^T$ will map integers $u$ and $v$ to the $n$th step of the Euclidean algorithm. These can be computed and multiplied together recursively.

Adapted to quantum circuits, these approaches require quantum matrix multiplication. We could find no efficient method to do this in-place, meaning that each recursive call would require a new set of auxiliary qubits to store the matrix output. This would quickly overwhelm our qubit budget. The base case of \cite{by2019} is nearly identical to our approach for a single round.

One of the primary advantages of \cite{by2019} is that the recursive process allows much of the arithmetic to be done with small integers which fit into the registers of classical CPUs. All the qubits in our model of a quantum computer are identical, so it has no caching or register issues. If quantum technologies arise with different kinds of qubits (perhaps a ``memory'' with higher coherence times but lower gate fidelity), then recursive GCD algorithms should be revisited. It is also possible that the specific structure of the matrices in this approach permit an easy, in-place multiplication circuit. We leave this to future work.

\subsection{Alternate curve representations}\label{app:curve_representations}
\subsubsection{Projective coordinates.}
Projective coordinates use equivalence classes $(X:Y:Z)$ of triples $(X,Y,Z)$ to represent an elliptic curve point, where $(X_1,Y_1,Z_1)\sim (X_2,Y_2,Z_2)$ iff there is some non-zero constant $c$ such that $X_1=cX_2$, $Y_1=cY_2$, and $Z_2=cZ_2$. 
These can be used with many different families of curves. Projective coordinates lend themselves to efficient, inversion-free arithmetic, which is appealing for classical computers.

Projective coordinates do not give a \emph{unique} representation of each point, which Shor's algorithm requires to ensure history independence and thus proper interference of states in superposition. Dividing by the $Z$ coordinate produces a unique representation but requires an expensive division. It is an open problem
to provide a unique projective representation with division-free arithmetic.

Another issue is that the classical elliptic curve formulas, naively adapted to quantum circuits, operate out-of-place. An out-of-place addition circuit is easy to adapt into an in-place addition circuit. If we can construct a circuit $U_{+Q}$ to add a point $Q$, we can construct a circuit $U_{-Q}$, and we can construct an in-place point addition\iffullversion{} as shown in Figure \ref{fig:in-place-point-addition} up to a final swap of the two qubit registers\else{} by writing $P+Q$ into another register, then subtracting $Q$ from $P+Q$ to clear the input\fi. This doubles the cost of point addition.

\iffullversion
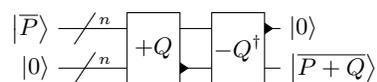
\begin{figure}
\centering
\input{quantum-circuits/Qpic/in-place-point}
\caption{An in-place point addition from out-of-place.}\label{fig:in-place-point-addition}
\end{figure}
\fi

This technique requires a unique representation. If $(P+Q)-Q$ does not have the same representation as $P$, we cannot cancel them out. Thus, for any current algorithm to compute addition with projective coordinates with cost $C$, we can transform it to a quantum-suitable in-place version with cost $2C+2D$, where $D$ is the cost of division. The division creates a unique representation.

\iffullversion
Some projective coordinate systems such as projective $x$-only Montgomery coordinates can only handle differential addition. In this case, we start with $P$, $Q$, and $Q-P$, and use $Q-P$ as a helper to compute $P+Q$. We did not find a method to clear the register with $P$ from the outputs $Q$, $Q-P$, and $P+Q$, so we are further restricted to curves that have direct addition formulas.\fi

According to the Explicit Formulas Database \cite{EFD}, the lowest-cost addition\iffullversion, possibly assuming $Z=1$ for the initial point,\fi{} uses 6 squares and/or multiplications. With the required reductions, the total cost is 12 squares/multiplications and 2 divisions, much higher than affine Weierstrass coordinates. Thus, we choose not to use projective coordinates in this work.

\iffullversion
\subsubsection{Elliptic curve point addition in affine Montgomery coordinates.}\label{app:affmont}
An alternative affine coordinate system is provided by elliptic curves in Montgomery form
\begin{equation}
by^2 = x^3 + ax^2 + x.
\end{equation}
Again, we represent points as pairs $(x,y)$ that satisfy the curve equation. 

Following the reference implementation for affine coordinate addition from \cite{SIKE}, we construct the circuit in Figure \ref{fig:mont-affine}. This uses one more register than affine Weierstrass coordinates, one more multiplication, and one more squaring. 

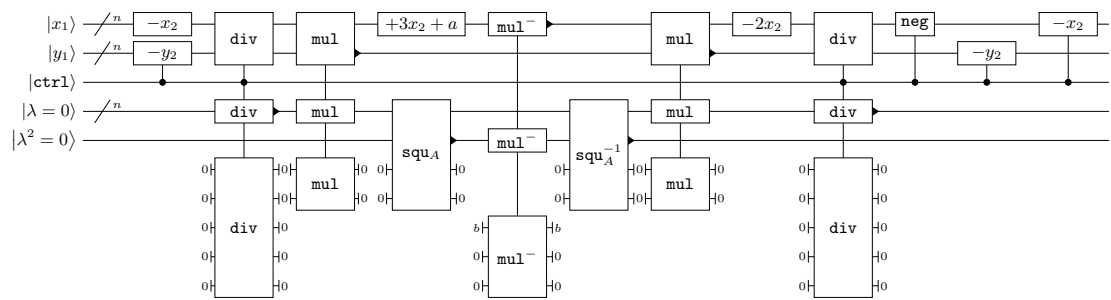
\begin{figure}
\resizebox{\textwidth}{!}{
\input{quantum-circuits/Qpic/mont-affine}
}
\caption{Constant point addition in affine Montgomery coordinates.}\label{fig:mont-affine}
\end{figure}

However, we note that the main source of extra costs is that, for a slope $\lambda$, there is a $b\lambda^3$ term. This requires 4 sequential multiplications/squarings. If we are lucky enough to have a curve where $b=1$, then we save space and operations by directly adding the $\lambda^2$ to the $x_1$ register. In fact, in this case, we can simply use the Weierstrass addition circuit, so long as we replace the $+3x_2$ operator with $+3x_2+a$. Thus, we have no reason to prefer affine Montgomery coordinates, and we will continue to use short Weierstrass form.
\fi

\subsection{Precomputation.}\label{app:precomputation}
Precomputed tables of certain powers of the base element can speed up exponentiations. The ``comb'' method is a standard technique used for elliptic curve scalar multiplication. To multiply a point $P$ by a scalar $k$, we divide $k$ into $k_1+2k_2+\cdots + 2^\ell k_\ell$ for some $\ell$, with the property that $k_j$ contains bits of $k$ in positions congruent to $j$ modulo $\ell$ (each $k_j$ looks like a comb of bits). We then precompute a table of all multiples of $P$ by scalars of the form $b_0+b_12^\ell+b_2 2^{2\ell}\dots$, with $b_i\in\{0,1\}$. By the definition of $k_j$, each $k_jP$ is a precomputed point in this table for all $j$. Thus, we can compute $kP$ by using $k_j$ to look up elements of the table, adding them to a running total, and doubling the running total. 

The advantage of the comb technique is that it saves precomputation. We only precompute one table and use it for the entire computation. Unfortunately for the quantum case, precomputation is essentially free because it is entirely classical, but look-ups are expensive. The comb technique does not reduce the number of table look-ups, since we must do a separate look-up for each index $k_j$. 

Further, efficient in-place point doubling is unlikely, since it implies efficient in-place point halving. Thus, doubling points in the comb would require some pebbling technique which would likely add significant depth or width costs.

\section{Modular division and addition}
For elliptic curve addition, we only need to divide integers and copy the result to a blank output, but other applications may wish to construct a circuit that, given registers containing $x$, $y$, and $z$, will compute $yx^{-1}+z$.

We might simply invert, multiply, and then add the output of the multiplication instead of copying. However, doubling the output to correct the pseudo-inverse while uncomputing will also multiply $z$ by a factor of $2^{2n-k}$. To correct for this, we can repeatedly halve $z$ during the \emph{forward} computation of the modular inverse. This means that while we compute the modular inverse, we control a modular halving of the register containing $z$ by the counter, which will halve $z$ exactly $2n-k$ times. Then we multiply the pseudo-inverse by $y$ and add the result to the register with $z$, producing the state
\iffullversion
\begin{equation}
\ket{x^{-1}2^{2n-k}\bmod p}\ket{y2^n\bmod p}\ket{z2^{2n-k}+x^{-1}y2^{2n-k}\bmod p}.
\end{equation}
\else
$\ket{x^{-1}2^{2n-k}\bmod p}\ket{y2^n\bmod p}\ket{z2^{2n-k}+x^{-1}y2^{2n-k}\bmod p}$.
\fi
From here, if we perform controlled modular doublings of the register containing $z$ as we uncompute the inversion circuit, this will correct both $z$ and the pseudo-inverse of $x$, producing the desired output.

\section{Analysis of windowed arithmetic}
A quantum look-up to $N$ elements requires $4N$ T-gates \cite{babbushetal2018}. To optimize window costs, we balance this cost against the operations we save. 

\subsubsection{Multiplication.}
Section~\ref{sec:multiplication_windowing} describes a single windowed multiplication round. For $n$-bit integers with window size $k$, repeating this round $\ceil{n/k}$ times performs the full multiplication. Since the quantum look-up will cost $4\cdot 2^k$ $T$ gates \cite{babbushetal2018} and uncontrolled $n+k$-bit addition costs $O(n+k)$ $T$ gates, we expect the optimal window size to be approximately $k=O(\lg n)$. The total multiplication cost is still $O(n^2)$ because we only window addition by $p$, not addition of the quantum register $y$. Compared to un-windowed add-and-halve multiplication, windowing should save a factor of roughly $\frac{1}{2}+ O(\frac{1}{\lg n})$. Similar reasoning suggests savings of $\frac{1}{2}+O(\frac{1}{\lg\lg n})$ in depth.

\iffullversion
Numerical estimates show a window size of $k=c_1\lg n + c_2$ optimizes $T$-count, where $0.68
\leq c_1\leq 0.75$ and $0.12\leq c_2\leq 0.78$,\else{}
Numerical estimates show a window size of $k\approx 0.7\lg n + 0.5$ optimizes $T$-count,\fi{}
and $1.97\lg\lg n-1.11$ optimizes $T$-depth. At the scale we estimate, this is only noticeable in the leading coefficient of the cost. We found a 22\% reduction in $T$-depth at $384$ bits, for example.

Windowing adds a significant cost of roughly $n+k$ auxiliary qubits, but the full elliptic curve point addition circuit has enough unused auxiliary qubits during any multiplication that this does not make a difference.

\subsubsection{Point addition.}
Windowing requires 2 extra registers as the cache to load the precomputed points. We use the components of the second point three times during point addition\iffullversion{} (see Figure \ref{fig:ECC_full})\fi. We could perform the look-up once and keep the values, increasing total circuit width by two registers. Alternatively, we can fit the look-ups within the existing space. \iffullversion{}As Figure~\ref{fig:ECC_full} shows, at\else{}At\fi{} every point where $x_2$ or $y_2$ are added, the circuit has spare auxiliary qubits available. Thus, we can perform the look-up, add the point to the quantum register, then uncompute the quantum look-up to free the qubits for the expensive modular division. This requires us six look-ups (including uncomputing) rather than just two, but uses no extra registers. 

With a window size of $\ell$, including sign bit, each look-up costs $4\cdot 2^{\ell - 1}$ $T$ gates and $T$-depth. The windowing saves us $\ell-1$ point additions. If point addition costs $\mathsf{A}$ T gates, we would expect $\ell\approx \lg(\mathsf{A}/24)$ to be the optimal value, leading to a factor $\ell$ reduction in T-gate cost.

\iffullversion
\section{Numerical results and interpolations}\label{app:asymptotic}
We also estimated the costs for the basic operations that comprise the elliptic curve addition. 

With \qsharp, we counted resources for arithmetic operations for all bit sizes from 4 up to 64, then for random bit sizes between 65 and 2048. Based on theoretical analysis and examining the best fit, we decided on the best model and then fit the data with linear regression. Tables~\ref{table:arithmetic} and \ref{table:alternate_arithmetic} show the results.

For modular arithmetic, we counted resources for all bit sizes from 4 to 64, then the common elliptic curve bit sizes 110, 160, 224, 256, 384, and 521. These were the only larger bit sizes we estimated due to the high computational cost of performing these estimates. With linear regression we estimated only the leading term of the cost growth. Table~\ref{table:modular_arithmetic} contains these costs. 

We found that addition with the DKRS adder had very different depths for numbers with an odd bitlength compared to an even bitlength. This behaviour is likely due to the structure of the binary tree for carry bit propagation. Odd bitlengths were much cheaper, so we opted to simply allocate an extra qubit to add two numbers of even bitlengths. 

We also noticed that addition and modular addition are substantially more expensive when controlled, while constant addition and modular doubling costs increase only negligibly. The T-depth of the uncontrolled DKRS adder increases monotonically with bitlength. The data for the controlled DKRS adder is much noisier. Hence, the apparent growth of the $\lg n$ term is slower than the controlled DKRS adder, which is nonsensical; this is likely just an artifact of the noise. 

For the final set of formulas for the entire cost of Shor's algorithm, we extrapolated the costs for a single point addition, then performed the same optimization as Section~\ref{sec:results} describes. Though the exact cost was not convex, it should be $O(n/\lg n)$ times the cost of a single addition, and indeed such a curve fit the data well ($r^2>0.98$ for the depth and gate costs).

\begin{table}
  \centering
  \begin{tabular}{| c | c | c | c | c | c | c |}
  \hline
  Circuit & Metric & Depth & Gates & \multicolumn{3}{c|}{Qubits} \\
  & & & & In/out & Auxiliary & Total\\ 
  \hline
  \hline
  \multicolumn{7}{| c |}{Fanin}\\
  \hline
  \multirow{2}{*}{Low W} & T & $13.50n-30.61$ & $56.0n-168$ & \multirow{2}{*}{$n$} &  \multirow{2}{*}{$2$} & \multirow{2}{*}{$n + 2$}\\
  & All & $40.5n-89.8$ & $130n-382$ &  &  & \\
  \hline
  \multirow{2}{*}{Low T/D} & T & $0.99\lg n+0.48$ & $4.00n-4.00$ & \multirow{2}{*}{$n$} &  \multirow{2}{*}{$n+1$} & \multirow{2}{*}{$2n+1$}\\
  & All & $6.98\lg n+8.38$ & $22.2n-16.2$ &  &  & \\
  \hline
  \hline
  \multicolumn{7}{| c |}{Addition}\\
  \hline
  \multirow{2}{*}{Low W} & T & $9.00n-4.00$ & $14.0n-7.0$ & \multirow{2}{*}{$2n$} &  \multirow{2}{*}{$1$} & \multirow{2}{*}{$2n+1$}\\
  & All & $31.0n-15.0$ & $47.0n-22.0$ &  &  & \\
  \hline
  \multirow{2}{*}{Low T} & T & $1.00n+0.00$ & $4.00n+0.00$ & \multirow{2}{*}{$2n$} &  \multirow{2}{*}{$n+2$} & \multirow{2}{*}{$3n+2$}\\
  & All & $12.0n+0.2$ & $30.2n+0.5$ &  &  & \\
  \hline
\multirow{2}{*}{Low D} & T & $3.86\lg n+1.16$ & $39.9 n-206$ & \multirow{2}{*}{$2n$} &  \multirow{2}{*}{$3n-6.8$}&  \multirow{2}{*}{$5n-6.8$} \\
& All & $28.8\lg n+14.5$ & $172 n-853$ &  &  &\\
\hline
\hline
  \multicolumn{7}{| c |}{Addition (Controlled)}\\
  \hline
  \multirow{2}{*}{Low W} & T & $14.00n+9.00$ & $21.0n+14.0$ & \multirow{2}{*}{$2n+1$} &  \multirow{2}{*}{$2$} & \multirow{2}{*}{$2n+3$}\\
  & All & $45.0n+13.8$ & $65.0n+36.0$ &  &  & \\
  \hline
  \multirow{2}{*}{Low T} & T & $11.0n-0.0$ & $18.0n+0.0$ & \multirow{2}{*}{$2n+1$} &  \multirow{2}{*}{$n+2$} & \multirow{2}{*}{$3n+3$}\\
  & All & $39.0n+3.0$ & $66.3n+4.7$ &  &  & \\
  \hline
\multirow{2}{*}{Low D} & T & $2.38\lg n+24.79$ & $53.9 n-191$ & \multirow{2}{*}{$2n+1$} &  \multirow{2}{*}{$3n-2.8$} &  \multirow{2}{*}{$5n-1.8$}\\
& All & $25.6\lg n+67.5$ & $222 n-815$ &  &  &\\
\hline
\hline
  \multicolumn{7}{| c |}{Comparator}\\
  \hline
  \multirow{2}{*}{Low W} & T & $9.00n-4.00$ & $14.0n-7.0$ & \multirow{2}{*}{$2n$} &  \multirow{2}{*}{$1$} & \multirow{2}{*}{$2n+1$}\\
  & All & $30.0n-12.0$ & $48.0n-22.0$ &  &  & \\
  \hline
  \multirow{2}{*}{Low T} & T & $1.00n+0.00$ & $4.00n+0.00$ & \multirow{2}{*}{$2n$} &  \multirow{2}{*}{$n+2$} & \multirow{2}{*}{$3n+2$}\\
  & All & $13.0n-1.4$ & $32.2n+1.0$ &  &  & \\
  \hline
  \multirow{2}{*}{Low D} & T & $3.75\lg n+1.27$ & $35.9 n-170$ & \multirow{2}{*}{$2n$} &  \multirow{2}{*}{$2n-1.4$} &  \multirow{2}{*}{$4n-1.4$} \\
& All & $27.7\lg n+3.9$ & $148 n-674$ &  & & \\
\hline
\hline
  \multicolumn{7}{| c |}{Comparator (Controlled)}\\
  \hline
  \multirow{2}{*}{Low W} & T & $9.00n+10.00$ & $14.0n+14.0$ & \multirow{2}{*}{$2n+1$} &  \multirow{2}{*}{$3$} & \multirow{2}{*}{$2n+4$}\\
  & All & $30.0n+30.0$ & $48.0n+36.0$ &  &  & \\
  \hline
  \multirow{2}{*}{Low T} & T & $1.00n+5.00$ & $4.00n+7.00$ & \multirow{2}{*}{$2n+1$} &  \multirow{2}{*}{$n+2$} & \multirow{2}{*}{$3n+3$}\\
  & All & $13.0n+12.7$ & $32.2n+13.3$ &  &  & \\
  \hline
  \multirow{2}{*}{Low D} & T & $3.08\lg n+12.7$ & $35.9 n-155$ & \multirow{2}{*}{$2n+1$} &  \multirow{2}{*}{$2n-1.4$} &  \multirow{2}{*}{$2n-0.4$} \\
& All & $25.5\lg n+27.5$ & $148 n-605$ &  &  &\\
\hline
\hline
  \multicolumn{7}{| c |}{Addition (No Carry)}\\
  \hline
  \multirow{2}{*}{Low W} & T & $9.00n-8.00$ & $14.0n-14.0$ & \multirow{2}{*}{$2n$} &  \multirow{2}{*}{$0$} &  \multirow{2}{*}{$2n$}\\
  & All & $31.0n-30.0$ & $47.0n-44.0$ &  &  &  \\
  \hline
  \multirow{2}{*}{Low T} & T & $1.00n+0.00$ & $4.00n+0.00$ & \multirow{2}{*}{$2n$} &  \multirow{2}{*}{$n+1$} &  \multirow{2}{*}{$3n+1$}\\
  & All & $12.0n-0.8$ & $30.2n-0.1$ &  &  &  \\
  \hline
 \multirow{2}{*}{Low D} & T & $4.03\lg n-0.49$ & $39.9 n-251$ & \multirow{2}{*}{$2n$} &  \multirow{2}{*}{$3n-10.1$}  &  \multirow{2}{*}{$5n-10.1$}\\
& All & $29.4\lg n+9.2$ & $172 n-1032$ &  &  &\\
\hline
\hline
  \multicolumn{7}{| c |}{Addition (No Carry, Controlled)}\\
  \hline
  \multirow{2}{*}{Low W} & T & $14.00n-9.00$ & $21.0n-14.0$ & \multirow{2}{*}{$2n+1$} &  \multirow{2}{*}{$0$} & \multirow{2}{*}{$2n+1$} \\
  & All & $45.0n-44.2$ & $65.0n-42.0$ &  &  &  \\
  \hline
  \multirow{2}{*}{Low T} & T & $11.0n-5.0$ & $18.0n-7.0$ & \multirow{2}{*}{$2n+1$} &  \multirow{2}{*}{$n+1$} & \multirow{2}{*}{$3n+2$} \\
  & All & $39.0n-12.0$ & $66.3n-22.9$ &  &  &  \\
  \hline
  \multirow{2}{*}{Low D} & T & $3.73\lg n+4.35$ & $53.9 n-258$ & \multirow{2}{*}{$2n+1$} &  \multirow{2}{*}{$3n-6.1$}  &  \multirow{2}{*}{$5n-5.1$}\\
& All & $29.2\lg n+24.8$ & $222 n-1054$ &  &  &\\
\hline
\hline
  \end{tabular}
  \caption{Basic arithmetic estimates.}\label{table:arithmetic}
\end{table}

\begin{table}
  \centering
  \begin{tabular}{| c | c | c | c | c | c | c |}
  \hline
  Circuit & Metric & Depth & Gates & \multicolumn{3}{c|}{Qubits} \\
  & & & & In/out & Auxiliary & Total\\ 
  \hline
  \hline
  \multicolumn{7}{| c |}{Constant Addition}\\
  \hline
\multirow{2}{*}{Low W} & T & $53.9n-307$ & $44n\lg n-7252$ & \multirow{2}{*}{$n$} &  \multirow{2}{*}{$1.6$}  &  \multirow{2}{*}{$n+1.6$} \\
& All & $176n-767$ & $122n\lg n-18983$ &  &  &\\
\hline
\multirow{2}{*}{Low T} & T & $1.00n+0.00$ & $4.00n+0.00$ & \multirow{2}{*}{$n$} &  \multirow{2}{*}{$2n+1$} & \multirow{2}{*}{$3n+1$} \\
  & All & $12.0n+1.7$ & $28.3n+80.7$ &  &  &  \\
  \hline
\multirow{2}{*}{Low D} & T & $4.03\lg n-1.49$ & $35.9 n-247$ & \multirow{2}{*}{$n$} &  \multirow{2}{*}{$3n-10.1$} &  \multirow{2}{*}{$4n-10.1$} \\
& All & $29.2\lg n+1.7$ & $147 n-940$ &  & & \\
\hline
\hline
  \multicolumn{7}{| c |}{Constant Addition (Controlled)}\\
  \hline
\multirow{2}{*}{Low W} & T & $19n\lg n-1870$ & $45n\lg n-6120$ & \multirow{2}{*}{$n+1$} &  \multirow{2}{*}{$0$} &  \multirow{2}{*}{$n+1$} \\
& All & $62n\lg n-5721$ & $126n\lg n-16076$ &  & & \\
\hline
 \multirow{2}{*}{Low T} & T & $1.00n+0.00$ & $4.00n+0.00$ & \multirow{2}{*}{$n+1$} &  \multirow{2}{*}{$2n+1$} & \multirow{2}{*}{$3n+2$} \\
  & All & $12.0n+3.8$ & $28.3n+87.7$ &  &  &  \\
  \hline
\multirow{2}{*}{Low D} & T & $4.03\lg n-0.49$ & $42.9 n-254$ & \multirow{2}{*}{$n+1$} &  \multirow{2}{*}{$3n-6.1$} &  \multirow{2}{*}{$4n-5.1$} \\
& All & $29.2\lg n+9.6$ & $173 n-950$ &  &  &\\
\hline
\hline
 \multicolumn{7}{| c |}{Modular Addition}\\
  \hline
  \multirow{2}{*}{Low W} & T & $30.8n\lg n-168$ & $87.0n\lg n-2433$ & \multirow{2}{*}{$2n$} &  \multirow{2}{*}{$1$} &  \multirow{2}{*}{$2n+1$}\\
  & All & $101n\lg n-779$ & $261n\lg n-8827$ &  &  &  \\ \hline
  \multirow{2}{*}{Low T} & T & $4.00n-2.00$ & $16.0n+4.0$ & \multirow{2}{*}{$2n$} &  \multirow{2}{*}{$2n+4$} &  \multirow{2}{*}{$4n+4$}\\
  & All & $49.0n-23.1$ & $111n+147$ &  &  &  \\ 
  \hline
\multirow{2}{*}{Low D} & T & $15.6\lg n+1.1$ & $155n-843$ & \multirow{2}{*}{$2n$} &  \multirow{2}{*}{$3n-3.7$}  &  \multirow{2}{*}{$5n-3.7$} \\
& All & $111{\lg n}+29$ & $633n-3268$ &  & & \\
\hline
\hline
  \multicolumn{7}{| c |}{Modular Addition (Controlled)}\\
  \hline
  \multirow{2}{*}{Low W} & T & $31.5n\lg n-210$ & $87.8n\lg n-2280$ & \multirow{2}{*}{$2n+1$} &  \multirow{2}{*}{$2$} & \multirow{2}{*}{$2n+3$} \\
  & All & $102n\lg n-523$ & $263n\lg n-8434$ &  &  &  \\ \hline
  \multirow{2}{*}{Low T} & T & $14.0n+4.0$ & $30.0n+11.0$ & \multirow{2}{*}{$2n+1$} &  \multirow{2}{*}{$2n+4$} &  \multirow{2}{*}{$4n+5$}\\
  & All & $76.0n+0.5$ & $147n+164$ &  &  &  \\ \hline
\multirow{2}{*}{Low D} & T & $13.4\lg n+36.2$ & $169n-812$ & \multirow{2}{*}{$2n+1$} &  \multirow{2}{*}{$3n-1.8$}  &  \multirow{2}{*}{$5n-0.8$}\\
& All & $106{\lg n}+102$ & $683n-3179$ &  &  &\\
\hline
  \hline
  \multicolumn{7}{| c |}{Modular Doubling}\\
  \hline
  \multirow{2}{*}{Low W} & T & $28.3n{\lg n}-484$ & $83.9n\lg n-2526$ & \multirow{2}{*}{$n$} &  \multirow{2}{*}{$2.5$} & \multirow{2}{*}{$n+2.5$} \\
  & All & $92.3n\lg n-1382$ & $252n\lg n-9314$ &  &  &  \\ \hline
  \multirow{2}{*}{Low T} & T & $2.00n+10.00$ & $15.0n+4.0$ & \multirow{2}{*}{$n$} &  \multirow{2}{*}{$2n+4$} &  \multirow{2}{*}{$3n+4$}\\
  & All & $24.0n+32.8$ & $79.8n+165.4$ &  &  &  \\ \hline
 \multirow{2}{*}{Low D} & T & $7.98{\lg n}+8.75$ & $85.8n-466$ & \multirow{2}{*}{$n$} &  \multirow{2}{*}{$3n-3.7$} &  \multirow{2}{*}{$4n-3.7$}\\
& All & $58.0{\lg n}+50.9$ & $343n-1757$ &  &  &\\
\hline
 \hline
  \multicolumn{7}{| c |}{Modular Doubling (Controlled)}\\
  \hline
  \multirow{2}{*}{Low W} & T & $28.9n{\lg n}-416$ & $84.7n\lg n-2451$ & \multirow{2}{*}{$n+1$} &  \multirow{2}{*}{$1.5$} & \multirow{2}{*}{$n+2.5$} \\
  & All & $94.0n\lg n-1185$ & $254n\lg n-9102$ &  &  &  \\ \hline
  \multirow{2}{*}{Low T} & T & $2.00n+14.0$ & $15.0n+11.0$ & \multirow{2}{*}{$n+1$} &  \multirow{2}{*}{$2n+5$} & \multirow{2}{*}{$3n+6$} \\
  & All & $24.0n+51.9$ & $79.9n+185$ &  &  &  \\ \hline
 \multirow{2}{*}{Low D} & T & $7.98{\lg n}+13.8$ & $85.8n-459$ & \multirow{2}{*}{$n+1$} &  \multirow{2}{*}{$3n-3.7$} &  \multirow{2}{*}{$4n-2.7$}  \\
& All & $60.0{\lg n}+61.8$ & $343n-1723$ &  &  &\\
\hline
 \hline
  \end{tabular}
  \caption{Alternate arithmetic estimates.}\label{table:alternate_arithmetic}
\end{table}

\begin{table}
  \centering
  \begin{tabular}{| c | c | c | c | c | c | c |}
  \hline
  Circuit & Metric & Depth & Gates & \multicolumn{3}{c|}{Qubits} \\
  & & & & In/out & Auxiliary & Total\\ 
  \hline
  \hline
  \multicolumn{7}{| c |}{Modular Squaring}\\
  \hline
\multirow{2}{*}{Low W} & T & $3.54n^2\lg n+1.32\cdot 2^{15}$ & $5.62n^2\lg n+1.24\cdot 2^{16}$ & \multirow{2}{*}{$2n$} &  \multirow{2}{*}{$3n+15.1$} &  \multirow{2}{*}{$5n+15.1$} \\
& All & $11.5n^2\lg n+1.09\cdot 2^{17}$ & $17.3n^2\lg n+1.94\cdot 2^{17}$ &  &  &\\
\hline
\multirow{2}{*}{Low T} & T & $21.4n^2+1.43\cdot 2^{10}$ & $37.8n^2+1.37\cdot 2^{12}$ & \multirow{2}{*}{$2n$} &  \multirow{2}{*}{$4n+12.3$} &  \multirow{2}{*}{$6n+12.3$} \\
& All & $75.7n^2+1.46\cdot 2^{13}$ & $143n^2+1.74\cdot 2^{15}$ &  &  &\\
\hline
\multirow{2}{*}{Low D} & T & $11.5n\lg n+1.88\cdot 2^{9}$ & $124n^2+1.29\cdot 2^{12}$ & \multirow{2}{*}{$2n$} &  \multirow{2}{*}{$6n+11.6$} &  \multirow{2}{*}{$8n+11.6$} \\
& All & $86.4n\lg n+1.40\cdot 2^{11}$ & $509n^2+1.56\cdot 2^{15}$ &  &  &\\
\hline
\hline
  \multicolumn{7}{| c |}{Modular Multiplication}\\
  \hline
\multirow{2}{*}{Low W} & T & $3.54n^2\lg n+1.35\cdot 2^{15}$ & $5.62n^2\lg n+1.28\cdot 2^{16}$ & \multirow{2}{*}{$3n$} &  \multirow{2}{*}{$3n+15.0$} &  \multirow{2}{*}{$6n+15.0$} \\
& All & $11.5n^2\lg n+1.12\cdot 2^{17}$ & $17.3n^2\lg n+2.00\cdot 2^{17}$ &  &  &\\
\hline
\multirow{2}{*}{Low T} & T & $21.4n^2+1.35\cdot 2^{10}$ & $37.8n^2+1.40\cdot 2^{12}$ & \multirow{2}{*}{$3n$} &  \multirow{2}{*}{$4n+12.2$} &  \multirow{2}{*}{$7n+12.2$} \\
& All & $75.7n^2+1.45\cdot 2^{13}$ & $143n^2+1.73\cdot 2^{15}$ &  &  &\\
\hline
\multirow{2}{*}{Low D} & T & $11.4n\lg n+1.84\cdot 2^{9}$ & $124n^2+1.29\cdot 2^{12}$ & \multirow{2}{*}{$3n$} &  \multirow{2}{*}{$6n+11.6$} &  \multirow{2}{*}{$9n+11.6$} \\
& All & $84.8n\lg n+1.42\cdot 2^{11}$ & $509n^2+1.54\cdot 2^{15}$ &  &  &\\
\hline
\hline
  \multicolumn{7}{| c |}{Modular Inversion}\\
  \hline
\multirow{2}{*}{Low W} & T & $72n^2\lg n+1.76\cdot 2^{16}$ & $236n^2\lg n+1.30\cdot 2^{18}$ & \multirow{2}{*}{$2n$} &  \multirow{2}{*}{$5n+1\floor{\lg n} +11$} &  \multirow{2}{*}{$7n+1\floor{\lg n} +11$}  \\
& All & $233n^2\lg n+1.44\cdot 2^{18}$ & $703n^2\lg n+1.49\cdot 2^{19}$ &  &  & \\ 
\hline
\multirow{2}{*}{Low T} & T & $162n^2+1.09\cdot 2^{12}$ & $496n^2+1.70\cdot 2^{14}$ & \multirow{2}{*}{$2n$} &  \multirow{2}{*}{$7n+1\floor{\lg n} +11$} &  \multirow{2}{*}{$9n+1\floor{\lg n} +11$}  \\
& All & $592n^2+1.42\cdot 2^{14}$ & $1789n^2+1.96\cdot 2^{16}$ &  &  & \\ 
\hline
\multirow{2}{*}{Low D} & T & $84.4n\lg n+1.10\cdot 2^{12}$ & $1062n^2-1.28\cdot 2^{16}$ & \multirow{2}{*}{$2n$} &  \multirow{2}{*}{$8n+12.8$} &  \multirow{2}{*}{$10n+12.8$} \\
& All & $529n\lg n+1.83\cdot 2^{13}$ & $4162n^2-1.28\cdot 2^{18}$ &  &  &\\
\hline
 \hline
  \multicolumn{7}{| c |}{Modular Division}\\
  \hline
\multirow{2}{*}{Low W} & T & $76n^2\lg n+1.26\cdot 2^{17}$ & $243n^2\lg n+1.66\cdot 2^{18}$ & \multirow{2}{*}{$3n$} &  \multirow{2}{*}{$5n+3.1\floor{\lg n} +9.7$} &  \multirow{2}{*}{$8n+3.1\floor{\lg n} +9.7$}  \\
& All & $245n^2\lg n+1.02\cdot 2^{19}$ & $722n^2\lg n+1.03\cdot 2^{20}$ &  &  & \\ 
\hline
\multirow{2}{*}{Low T} & T & $184n^2+1.46\cdot 2^{12}$ & $534n^2+1.03\cdot 2^{15}$ & \multirow{2}{*}{$2n$} &  \multirow{2}{*}{$8n+1\floor{\lg n} +11$} &  \multirow{2}{*}{$10n+1\floor{\lg n} +11$}  \\
& All & $667n^2+1.12\cdot 2^{15}$ & $1932n^2+1.43\cdot 2^{17}$ &  &  & \\ 
\hline
\multirow{2}{*}{Low D} & T & $95.8n\lg n+1.35\cdot 2^{12}$ & $1186n^2-1.13\cdot 2^{16}$ & \multirow{2}{*}{$2n$} &  \multirow{2}{*}{$9n+18.8$} &  \multirow{2}{*}{$11n+18.8$} \\
& All & $614n\lg n+1.12\cdot 2^{14}$ & $4672n^2-1.01\cdot 2^{18}$ &  &  &\\
\hline
\hline
  \multicolumn{7}{| c |}{Signed Windowed Elliptic Curve Point Addition (window size 8)}\\
  \hline
\multirow{2}{*}{Low W} & T & $144n^2\lg n+1.19\cdot 2^{19}$ & $503n^2\lg n+1.26\cdot 2^{20}$ & \multirow{2}{*}{$2n$} &  \multirow{2}{*}{$6n+3.8\floor{\lg n} +17.1$} &  \multirow{2}{*}{$8n+3.8\floor{\lg n} +17.1$}  \\
& All & $465n^2\lg n+1.98\cdot 2^{20}$ & $1423n^2\lg n+1.99\cdot 2^{21}$ &  &  & \\ 
\hline
\multirow{2}{*}{Low T} & T & $432n^2+1.07\cdot 2^{14}$ & $1182n^2+1.41\cdot 2^{16}$ & \multirow{2}{*}{$2n$} &  \multirow{2}{*}{$8n+1.5\floor{\lg n} +18.9$} &  \multirow{2}{*}{$10n+1.5\floor{\lg n} +18.9$}  \\
& All & $1562n^2+1.85\cdot 2^{16}$ & $4306n^2+1.31\cdot 2^{19}$ &  &  & \\ 
\hline
\multirow{2}{*}{Low D} & T & $226n\lg n+1.77\cdot 2^{13}$ & $2746n^2-1.31\cdot 2^{16}$ & \multirow{2}{*}{$2n$} &  \multirow{2}{*}{$9n+28.6$} &  \multirow{2}{*}{$11n+28.6$} \\
& All & $1485n\lg n+1.60\cdot 2^{15}$ & $10891n^2-1.39\cdot 2^{16}$ &  &  &\\
\hline
\hline
  \multicolumn{7}{| c |}{Shor's algorithm (with signed windowed point addition)}\\
  \hline
\multirow{2}{*}{Low W} & T & $120n^3-1.67\cdot 2^{22}$ & $436n^3-1.05\cdot 2^{26}$ & \multirow{2}{*}{$2n$} &  \multirow{2}{*}{$6n+10.2\floor{\lg n} -1.0$} &  \multirow{2}{*}{$8n+10.2\floor{\lg n} -1.0$}  \\
& All & $509n^3-1.84\cdot 2^{27}$ & $2800n^3 -1.08\cdot 2^{31}$ &  &  & \\ 
\hline
\multirow{2}{*}{Low T} & T & $389n^3/\lg n-1.70\cdot 2^{22}$ & $1115n^3/\lg n-1.08\cdot 2^{24}$ & \multirow{2}{*}{$2n$} &  \multirow{2}{*}{$8n+7.4\floor{\lg n} +1.3$} &  \multirow{2}{*}{$10n+7.4\floor{\lg n} +1.3$}  \\
& All & $1701n^3/\lg n-1.23\cdot 2^{24}$ & $6262n^3/\lg n -1.72\cdot 2^{24}$ &  &  & \\ 
\hline
\multirow{2}{*}{Low D} & T & $285n^2-1.54\cdot 2^{17}$ & $3120n^3/\lg n-1.49\cdot 2^{27}$ & \multirow{2}{*}{$2n$} &  \multirow{2}{*}{$9n+3.9\floor{\lg n} +16.5$}  &  \multirow{2}{*}{$11n+3.9\floor{\lg n} +16.5$}  \\
& All & $2523n^2+1.10\cdot 2^{20}$ & $12478n^3/\lg n -1.25\cdot 2^{29}$ &  &  & \\
\hline
\hline
  \end{tabular}
  \caption{Estimates for non-linear modular arithmetic, elliptic curve point addition and Shor's algorithm.}\label{table:modular_arithmetic}
\end{table}
\fi

\section{Automatic compilation for aggressive T-count and T-depth reduction}\label{app:autocompile}

In this section, we motivate automatic compilation methods to
drastically reduce the $T$-count and the $T$-depth if we allow a
significant increase in circuit width.

The modular multiplication followed by an addition is one of the most costly operations in the
overall algorithm.  It is implemented as a unitary
$U : |x\rangle|y\rangle|z\rangle|0\rangle \mapsto |x\rangle|y\rangle|(xy + z) \bmod p\rangle|0\rangle$
that adds the result of the multiplication of two numbers $x$ and $y$
onto a third number $z$, all in Montgomery form with bit-width $n$ and
modulus $p$.  We apply the following procedure to automatically obtain
a quantum circuit for this operation:

\begin{figure}[t]
  \centering
  \footnotesize
  \input{quantum-circuits/Qpic/modmult.tikz}
  \caption{Quantum circuit that implements
    $xy + z \bmod p$, using out-of-place
    constructions for modular multiplication, modular
    addition, and modular subtraction.}
  \label{fig:modmult}
\end{figure}
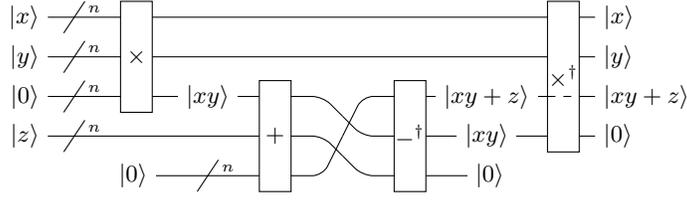

\begin{enumerate}
\item We generate logic networks over the gate basis
  $\{\mathrm{AND}, \mathrm{XOR}, \mathrm{INV}\}$, called
  Xor-And-inverter Graphs (XAGs), for the functions $xy \bmod p$,
  $(x + y) \bmod p$, and $(x - y) \bmod p$, where $x$ and $y$ are
  integers in Montgomery form.
\item We apply the logic optimization method described
  in~\cite{tsam19} to minimize the number of AND gates in the XAGs.
\item The optimized XAGs are then translated into out-of-place quantum
  circuits using the method in~\cite{msc+19}, which requires 4 $T$
  gates for each AND gate in the XAG.  Optimizing these circuits
  for depth requires roughly 2 qubits for each AND gate in the
  XAG, by using the AND gate construction from
  Section~\ref{sec:components}.
\item The automatically generated unitaries are composed as described
  in Figure~\ref{fig:modmult}, which uses \iffullversion{}the technique shown in
  Figure~\ref{fig:in-place-point-addition}\else{}a technique similar to that described in Section~\ref{app:curve_representations}\fi{} to turn the out-of-place
  addition and subtraction into an in-place addition.
\end{enumerate}

\begin{table}
  \caption{Comparison of resource costs between a manual and automatic construction to implement $|xy + z \bmod p\rangle$.}
  \label{tab:modmult-results}
  \centering
  \begin{tabular}{|c|r|r|r|r|r|r|}
    \hline
    & \multicolumn{3}{|c|}{Manual construction} & \multicolumn{3}{|c|}{Automatic construction} \\\cline{2-7}
    Bit-width & $T$-count & $T$-depth & Width & $T$-count & $T$-depth & Width \\\hline
    256 &  8,176,739 &  50,253 & 2,319 & 1,576,296 & 1,542 &   394,588 \\\hline
    384 & 18,322,671 &  76,125 & 3,470 & 3,550,552 & 2,310 &   888,408 \\\hline
    521 & 33,751,240 & 137,183 & 4,702 & 6,535,384 & 3,132 & 1,634,890 \\\hline
  \end{tabular}
\end{table}

Table~\ref{tab:modmult-results} lists the resource costs in terms of
$T$-count, $T$-depth, and circuit width, for both the manual construction
and the automatic construction.  Several factors of reduction in
$T$-count and $T$-depth are possible, while the increase in the number of
qubits is significant.  However, such a design point can be of high
interest, in particular when combined with automatic quantum memory
strategies, e.g., pebbling~\cite{msr+19}, that can find intermediate
trade-off points that lie in between the manual and automatic
construction.

%%% Local Variables:
%%% mode: latex
%%% TeX-master: "ECDLP"
%%% End:

%% file: quantum-circuits/Qpic/in-place-point
PREAMBLE \providecommand{\ket}[1]{\left\vert #1\right\rangle}
PREAMBLE \providecommand{\gtxt}[1]{\texttt{#1}}

point W \ket{\overline{P}} \ket{0}
PQ W \ket{0} \ket{\overline{P+Q}}

point / n
PQ / n

point PQ G $+Q$ width=20

point PQ G $-Q$ width=20

%% file: quantum-circuits/Qpic/mont-affine
PREAMBLE \providecommand{\ket}[1]{\left\vert #1\right\rangle}
PREAMBLE \providecommand{\gtxt}[1]{\texttt{#1}}

x1 W \ket{x_1}
y1 W \ket{y_1}
ctrl W \ket{\gtxt{ctrl}}
m W \ket{\lambda=0}
m2 W \ket{\lambda^2=0}
t0 W
t1 W
t2 W
t3 W
t4 W

x1 / n
y1 / n
m / n

x1 G $-x_2$ width=30
y1 G $-y_2$ width=30 ctrl
x1 y1 G \gtxt{div} m G \gtxt{div} t0 t1 t2 t3 t4 G \gtxt{div} ctrl width=30
x1 y1 G \gtxt{mul} m G \gtxt{mul} t0 t1 G \gtxt{mul} width=30
TOUCH
x1 G $+3x_2+a$ width=45
m m2 t0 t1 G \gtxt{squ}$_A$ width=30
t2 G $b$
x1 G \gtxt{mul}$^-$ m2 G \gtxt{mul}$^-$ t2 t3 t4 G \gtxt{mul}$^-$ width=30
t2 G $b$
m m2 t0 t1 G \gtxt{squ}$_A^{-1}$ width=30
x1 y1 m t0 t1 G \gtxt{mul} width=30
x1 G $-2x_2$ width=30
x1 y1 G \gtxt{div} m G \gtxt{div} t0 t1 t2 t3 t4 G \gtxt{div} ctrl width=30
x1 G \gtxt{neg} width=20 ctrl
y1 G $-y_2$ width=30 ctrl
x1 G $-x_2$ width=30 ctrl

%% file: quantum-circuits/Qpic/modmult.tikz
\tikzpicture[scale=1.000000,x=1pt,y=1pt]
\filldraw[color=white] (0.000000, -7.500000) rectangle (207.000000, 67.500000);
% Drawing wires
% Line 1: x  W |x\rangle |x\rangle
\draw[color=black] (0.000000,60.000000) -- (207.000000,60.000000);
\draw[color=black] (0.000000,60.000000) node[left] {$|x\rangle$};
% Line 2: y  W |y\rangle |y\rangle
\draw[color=black] (0.000000,45.000000) -- (207.000000,45.000000);
\draw[color=black] (0.000000,45.000000) node[left] {$|y\rangle$};
% Line 3: c0 W |0\rangle |0\rangle
\draw[color=black,rounded corners=4.000000pt] (0.000000,30.000000) -- (104.000000,30.000000) -- (111.500000,22.500000);
\draw[color=black,rounded corners=4.000000pt] (111.500000,22.500000) -- (119.000000,15.000000) -- (207.000000,15.000000);
\draw[color=black] (0.000000,30.000000) node[left] {$|0\rangle$};
% Line 4: z  W |z\rangle |0\rangle
\draw[color=black,rounded corners=4.000000pt] (0.000000,15.000000) -- (104.000000,15.000000) -- (111.500000,7.500000);
\draw[color=black,rounded corners=4.000000pt] (111.500000,7.500000) -- (119.000000,0.000000) -- (166.000000,0.000000);
\draw[color=black] (0.000000,15.000000) node[left] {$|z\rangle$};
% Line 5: c1 W |0\rangle {|xy + z\rangle}
\draw[color=black,rounded corners=4.000000pt] (33.500000,0.000000) -- (104.000000,0.000000) -- (111.500000,15.000000);
\draw[color=black,rounded corners=4.000000pt] (111.500000,15.000000) -- (119.000000,30.000000) -- (207.000000,30.000000);
% Done with wires; drawing gates
% Line 7: x  / n
\draw (6.000000, 54.000000) -- (14.000000, 66.000000);
\draw (12.000000, 63.000000) node[right] {$\scriptstyle{n}$};
% Line 8: y  / n
\draw (6.000000, 39.000000) -- (14.000000, 51.000000);
\draw (12.000000, 48.000000) node[right] {$\scriptstyle{n}$};
% Line 9: c0 / n
\draw (6.000000, 24.000000) -- (14.000000, 36.000000);
\draw (12.000000, 33.000000) node[right] {$\scriptstyle{n}$};
% Line 10: z  / n
\draw (6.000000, 9.000000) -- (14.000000, 21.000000);
\draw (12.000000, 18.000000) node[right] {$\scriptstyle{n}$};
% Line 12: x y c0 G {$\times$}
\draw (33.500000,60.000000) -- (33.500000,30.000000);
\scope
\draw[fill=white] (33.500000, 45.000000) +(-45.000000:8.485281pt and 29.698485pt) -- +(45.000000:8.485281pt and 29.698485pt) -- +(135.000000:8.485281pt and 29.698485pt) -- +(225.000000:8.485281pt and 29.698485pt) -- cycle;
\clip (33.500000, 45.000000) +(-45.000000:8.485281pt and 29.698485pt) -- +(45.000000:8.485281pt and 29.698485pt) -- +(135.000000:8.485281pt and 29.698485pt) -- +(225.000000:8.485281pt and 29.698485pt) -- cycle;
\draw (33.500000, 45.000000) node {{$\times$}};
\endscope
% Line 13: c1 START
\draw[color=black] (41.000000,0.000000) node[fill=white,left,minimum height=15.000000pt,minimum width=15.000000pt,inner sep=0pt] {\phantom{$|0\rangle$}};
\draw[color=black] (41.000000,0.000000) node[left] {$|0\rangle$};
% Line 14: c0 LABEL |xy\rangle
\draw[color=black] (60.500000, 30.000000) node [fill=white] {$|xy\rangle$};
% Line 15: c1 / n
\draw (56.500000, -6.000000) -- (64.500000, 6.000000);
\draw (62.500000, 3.000000) node[right] {$\scriptstyle{n}$};
% Line 16: c0 z c1 G {$+$}
\draw (86.000000,30.000000) -- (86.000000,0.000000);
\scope
\draw[fill=white] (86.000000, 15.000000) +(-45.000000:8.485281pt and 29.698485pt) -- +(45.000000:8.485281pt and 29.698485pt) -- +(135.000000:8.485281pt and 29.698485pt) -- +(225.000000:8.485281pt and 29.698485pt) -- cycle;
\clip (86.000000, 15.000000) +(-45.000000:8.485281pt and 29.698485pt) -- +(45.000000:8.485281pt and 29.698485pt) -- +(135.000000:8.485281pt and 29.698485pt) -- +(225.000000:8.485281pt and 29.698485pt) -- cycle;
\draw (86.000000, 15.000000) node {{$+$}};
\endscope
% Line 17: c1 c0 z PERMUTE
% Line 18: c0 z c1 G {$-^\dagger$}
\draw (137.000000,30.000000) -- (137.000000,0.000000);
\scope
\draw[fill=white] (137.000000, 15.000000) +(-45.000000:8.485281pt and 29.698485pt) -- +(45.000000:8.485281pt and 29.698485pt) -- +(135.000000:8.485281pt and 29.698485pt) -- +(225.000000:8.485281pt and 29.698485pt) -- cycle;
\clip (137.000000, 15.000000) +(-45.000000:8.485281pt and 29.698485pt) -- +(45.000000:8.485281pt and 29.698485pt) -- +(135.000000:8.485281pt and 29.698485pt) -- +(225.000000:8.485281pt and 29.698485pt) -- cycle;
\draw (137.000000, 15.000000) node {{$-^\dagger$}};
\endscope
% Line 19: z END
\draw[color=black] (158.500000,0.000000) node[fill=white,right,minimum height=15.000000pt,minimum width=15.000000pt,inner sep=0pt] {\phantom{$|0\rangle$}};
\draw[color=black] (158.500000,0.000000) node[right] {$|0\rangle$};
% Line 20: c0 c1 LABEL |xy\rangle |xy+z\rangle width=22
\draw[color=black] (166.000000, 15.000000) node [fill=white] {$|xy\rangle$};
\draw[color=black] (166.000000, 30.000000) node [fill=white] {$|xy+z\rangle$};
% Line 21: x y c0 G {$\times^\dagger$}
\draw (195.000000,60.000000) -- (195.000000,15.000000);
\scope
\draw[fill=white] (195.000000, 37.500000) +(-45.000000:8.485281pt and 40.305087pt) -- +(45.000000:8.485281pt and 40.305087pt) -- +(135.000000:8.485281pt and 40.305087pt) -- +(225.000000:8.485281pt and 40.305087pt) -- cycle;
\clip (195.000000, 37.500000) +(-45.000000:8.485281pt and 40.305087pt) -- +(45.000000:8.485281pt and 40.305087pt) -- +(135.000000:8.485281pt and 40.305087pt) -- +(225.000000:8.485281pt and 40.305087pt) -- cycle;
\draw (195.000000, 37.500000) node {{$\times^\dagger$}};
\endscope
\draw[color=black,dashed] (189.000000, 30.000000) -- (201.000000, 30.000000);
% Done with gates; drawing ending labels
\draw[color=black] (207.000000,60.000000) node[right] {$|x\rangle$};
\draw[color=black] (207.000000,45.000000) node[right] {$|y\rangle$};
\draw[color=black] (207.000000,15.000000) node[right] {$|0\rangle$};
\draw[color=black] (207.000000,30.000000) node[right] {${|xy + z\rangle}$};
% Done with ending labels; drawing cut lines and comments
% Done with comments
\endtikzpicture